\documentclass[11pt]{article}
\usepackage{amssymb,latexsym, amsmath}
\usepackage{amscd}

\makeatletter \@addtoreset{figure}{section}
\def\thefigure{\thesection.\@arabic\c@figure}
\def\fps@figure{h, t}
\@addtoreset{table}{bsection}
\def\thetable{\thesection.\@arabic\c@table}
\def\fps@table{h, t}
\@addtoreset{equation}{section}

\newtheorem{thm}{Theorem}[section]
\newtheorem{prop}[thm]{Proposition}
\newtheorem{lem}[thm]{Lemma}
\newtheorem{cor}[thm]{Corollary}
\newtheorem{remark}{Remark}[section]

\newcommand{\h}{\mathfrak{ h}}
\newcommand{\g}{\mathfrak{ g}}
\newcommand{\D}{\mathfrak{ d}}
\newcommand{\N}{\mathcal{ N}}
\DeclareMathOperator{\pr}{pr}
\DeclareMathOperator{\Span}{span}

\begin{document}
\title{Nonholonomic  LR Systems as  Generalized Chaplygin Systems
with an Invariant Measure and Geodesic Flows on Homogeneous Spaces
\footnote{AMS Subject Classification 37J60, 37J35, 70H06, 70H45}}
\author{Yuri N. Fedorov \\
 Department of Mathematics and Mechanics
 \\ Moscow Lomonosov University, Moscow, 119 899, Russia \\
e-mail: fedorov@mech.math.msu.su \\
and \\
 Department de Matem\`atica I, \\
Universitat Politecnica de Catalunya, \\
Barcelona, E-08028 Spain \\
e-mail: Yuri.Fedorov@upc.es \\
\and
Bo\v zidar Jovanovi\' c\\
Mathematical Institute, SANU \\
Kneza Mihaila 35, 11000, Belgrad, Serbia \\
e-mail: bozaj@mi.sanu.ac.yu}
\maketitle

\abstract    We consider a class of dynamical systems on a compact Lie group $G$ with a
left-invariant metric and right-invariant nonholonomic constraints (so called LR systems)
and show that, under a generic condition on the constraints, such systems can be regarded
as generalized Chaplygin systems on the principle bundle $G \to Q=G/H$, $H$ being a Lie subgroup.
In contrast to generic Chaplygin systems, the reductions of our LR systems onto the
homogeneous space $Q$ always possess an invariant measure.

We study the case $G=SO(n)$, when LR systems are
multidimensional generalizations of the Veselova problem of a nonholonomic rigid body motion
which admit a reduction to systems with an invariant measure on the (co)tangent bundle of
Stiefel varieties $V(k,n)$ as the corresponding homogeneous spaces.

For $k=1$ and a special choice of the left-invariant metric on $SO(n)$,
we prove that after a time substitution, the reduced system becomes an integrable Hamiltonian
system describing a geodesic flow on the unit sphere $S^{n-1}$. This provides a first example of
a nonholonomic system with more than two degrees of freedom for which the celebrated
Chaplygin reducibility theorem is applicable for any dimension.
In this case we also explicitly reconstruct the motion on the group $SO(n)$.

\endabstract

\tableofcontents

\section{Introduction}
In classical nonholonomic mechanics a special attention is given to Chaplygin systems whose
Lagrangian and constraints admit symmetries such that
after an appropriate reduction the equations of motion take the form
of unconstrained Lagrangian systems with some extra (nonholonomic) forces.
Excellent reviews of the history, various forms and geometric descriptions of the reduced systems,
as well as many relevant examples can be found in \cite{NeFu, Koi, BS, CdLMMdD, BKMM, Sumb}, see
also references therein.

Apparently, Appel \cite{Ap} was the first to propose time substitution
in order to eliminate these extra terms and to transform the reduced systems to a canonical
(Hamiltonian) form.  After that, Chaplygin \cite{Ch} realized this idea in his
reducing multiplier theory for nonholonomic systems with two degrees of freedom.

The key feature in Chaplyginïs approach is
the existence of an invariant measure of the reduced system, a rather strong property
which puts the system close to Hamiltonian ones.
For reduced generalized Chaplygin systems emerged from classical dynamics, this problem
was considered in \cite{Koi}.
Recently, necessary and sufficient conditions for the existence of such a measure
when the Lagrangian of the system is of a pure kinetic energy type are given in \cite{CCLM}.

On the other hand, numerous attempts to extend Chaplyginïs reducing multiplier theory to
systems with more than two degrees of freedom (even having an invariant measure)
were ineffective, since in this case several overdetermined conditions on the metric
and constraints are imposed (\cite{Efimov, Iliev}).
To our knowledge, until recently there were very few nontrivial examples
of multidimensional systems
that are reducible to a Hamiltonian form exactly by the Chaplygin procedure
(\cite{Moshchuk, Iliev, EKR}). We also quote the results of \cite{BM, BMK}, where some
clasical nonholonomic problems were reduced to Hamiltonian flows with respect to a nonlinear
Poisson bracket.

As an alternative, much effort has gone into the development of the
symplectic and Poisson view of reduced generalized Chaplygin systems.
In particular, in the case of Abelian symmetries,
Stanchenko \cite{St} showed that reduced systems can be represented in a
Hamilton-like form with respect to an almost symplectic 2-form $\varOmega$,
which is generally not closed. This observation was extended for generic
symmetries (see \cite{BS, KM, CCLM}). In this framework, the Chaplygin multiplier
is a function $f$ such that the form $f\varOmega$ is closed (see \cite{St, He, CCLM}).

The importance of the existence of an invariant measure for integrability of
nonholonomic systems was also indicated by Kozlov in \cite{Koz1, Koz2}, where various examples
were considered. In \cite{VeVe1, VeVe2},
Veselov and Veselova, inspired by classical problems of nonholonomic dynamics,
studied nonholonomic systems on unimodular
Lie groups with right-invariant nonintegrable constraints and a left-invariant metric
(so called LR systems),
and showed that they always possess an invariant measure whose density can be effectively
calculated. In particular, the motion of a rigid body  around a fixed point under a nonholonomic constraint (projection of the angular velocity to the fixed vector in space
is constant) is described  by an integrable LR system (\cite{VeVe1}).

Another  method of constructing non-Lagrangian (so called L+R)
systems with an invariant measure on  Lie groups was proposed in \cite{Fe1}.
The kinetic energy of such systems is given by a sum of left- and right-invariant metrics
on the group.  It appears that
some L+R systems have natural origins in classical nonholonomic mechanics.

For related problems concerning the integrability of nonholonomic
systems one can see \cite{BKMM, Koz1, FeKo2, Ze1, He, BaCu, Jo2, Jo3} and references therein.
Also, the existence of an invariant measure for
a class of nonholonomic systems with symmetries, which include the Chaplygin systems,
is recently studied in \cite{Bl_Z3}.

\paragraph{Contents of the paper.} We study several new geometric aspects of nonholonomic
LR systems on a compact Lie group $G$.
In Section 2 we show that a class of such systems can be naturally considered as
generalized Chaplygin systems on the principle bundle $G \to Q=G/H$, where
$H$ is a subgroup of $G$.  Such systems are reduced to non-Hamiltonian
equations on the cotangent bundle of the homogeneous space $Q$.
The latter are described  by a Lagrange-d'Alembert equation with
extra nonholonomic terms which are explicitly found.

In Section 3 we describe the invariant measure of the original and reduced LR systems.
If the homogeneous space is two-dimensional, then, by the Chaplygin reducibility theorem,
the existence of such a measure leads to a time substitution
such that our system becomes Hamiltonian.
On the other hand, we prove that if the reduced system is transformable in this way
to a Hamiltonian form for any dimension, then it must have an invariant measure whose density is
prescribed by the corresponding reducing multiplier.
We also show that the reduced LR system on $Q$ always possesses an invariant measure;
this does not necessarily holds for generic Chaplygin systems.

As a natural example of LR systems,
Section 4 describes the classical Veselova problem on the motion of
a rigid body with a nonholonomic constraint and some
of its integrable perturbations, as well as its relation to the Neumann system and
an integrable geodesic flow on the 2-dimensional sphere.

In Section 5 we consider multidimensional Veselova nonholonomic systems on the Lie
group $SO(n)$ characterized by various types of constraints and describe their invariant
measure. The constraints allow a reduction of these systems to non-Hamiltonian flows
with an invariant measure on the cotangent bundle of Stiefel varieties $V(r,n)$.

In Section 6 we concentrate on the case $r=1$, which corresponds to reduced flows
on the unit sphere $S^{n-1}$. We show that for a special choice of the inertia tensor
and after a time substitution, the flow reduces to a completely integrable geodesic
flow on the sphere. This provides a first example of
a nonholonomic system with more than two degrees of freedom for which the
celebrated Chaplygin reducibility theorem is applicable for any dimension.

Also, we prove that, under another time substitution, the multidimensional
Veselova nonholonomic system on $SO(n)$ reduces to the Neumann system on $S^{n-1}$.

In in last section, for the above integrable case, we explicitly solve the reconstruction problem:
given a trajectory of the reduced geodesic flow on $S^{n-1}$, to find
the corresponding nonholonomic motion on the group $SO(n)$. To perform this,
we use the remarkable relations between the Neumann system, the geodesic flow
on an $(n-1)$-dimensional ellipsoid, and the evolution of orthogonal frames associated
to the geodesics. It appears that the right-invariant distribution $D\subset T SO(n)$
is foliated
with invariant tori of generic dimension $n-1$ and the unreduced LR system is integrable.

\section{Generalized Chaplygin and LR systems on Lie groups}

Suppose we are given a nonholonomic Lagrangian system $(M,l,D)$ on the $n$--dimensional
configuration space $M$ with (local) coordinates $x$ and Lagrangian $l(x,\dot x)$ in the presence
of a $k$--dimensional distribution $D\subset TM$
describing kinematic constraints: a curve $x(t)$ is said to satisfy the
constraints if $\dot x(t)\in D_{x(t)}$ for all $t$.
The trajectory of the system $x(t)$ that satisfies the constraints
is a solution to the Lagrange-d'Alembert equation
\begin{equation}
\left( \frac{\partial l}{\partial x}
- \frac{d}{dt}\frac{\partial l}{\partial \dot x},\eta\right)=0,
\quad \mathrm{for\;all} \quad \eta\in D_x.  \label{1.1}
\end{equation}
Here $(\cdot,\cdot)$ denotes pairing between dual spaces.

Now assume that $M$ has a bundle structure $\pi: M\to Q$ with a base
manifold $Q$ and the map $\pi$ is a submersion, that is,
$T_x M= D_x\oplus V_x$ for all $x$. Here $V_x$ is the kernel
of $T_x\pi$ and  it is  called the vertical space at $x$.
Then the distribution $D$ can be seen as a collection of horizontal spaces
of the  {\it Ehresmann connection} associated with $\pi: M\to Q$.

Given a vector $X_x\in T_x M$, we have the decomposition $X_x=X_x^h+X_x^v$, where
$X_x^h\in D_x$, $X_x^v\in V_x$.
The {\it curvature} of the connection is the vertical valued two form $B$ on $M$
defined by
$$
B(X_x,Y_x)=-[\bar X_x^h,\bar Y_x^h]_x^v
$$
where $\bar X$ and $\bar Y$ are smooth vector fields on $M$ obtained by extending
$X_x$ and $Y_x$.

With the help of the Ehresmann connection, the equations of motion can be
put into the form (see \cite{BKMM})
\begin{equation}
\left( \frac{\partial l_c}{\partial x}
- \frac{d}{dt}\frac{\partial l_c}{\partial \dot x},\eta\right)=
\left(\frac{\partial l}{\partial \dot x},B(\dot x,\eta)\right),
\quad \mathrm{for\;all} \quad \eta\in D_x,  \label{1.1a}
\end{equation}
where $l_c(x,\dot x)=l(x,\dot x^h)$ is the constrained Lagrangian.

The form of equations (\ref{1.1a}) is very useful in the presence of
symmetries of the system. Namely,
suppose that the configuration space is a principal bundle $\pi: M\to Q=M/{\mathfrak G}$
with respect to the (left) action of a Lie group ${\mathfrak G}$,
and $D$ is a principal connection (i.e., $D$ is a $\mathfrak G$--invariant distribution).
Let the Lagrangian $l$ also be $\mathfrak G$--invariant.
Then  equations (\ref{1.1a}) are $\mathfrak G$-invariant and
induce a well defined reduced Lagrange-d'Alembert equation on the tangent bundle
$TQ=D/{\mathfrak G}$. The system $(M,l,D)$ is referred to as {\it a generalized Chaplygin system}
(see \cite {Koi, BKMM}).
\medskip

\paragraph{LR systems.}
Now let $M$ be a compact connected Lie group $G$ of dimension $n$ with local coordinates $g$,
and $\mathfrak{g}=T_{Id} G$ its Lie
algebra with comutator $[\, ,\,]$. Let $\langle \cdot,\cdot \rangle$ denote
either the $Ad_G$-invariant scalar
product on $\mathfrak{g}$ or the bi-invariant scalar product on $G$, and $ds^2_{\mathcal I}$
denote the left-invariant metric on $G$ given by the nondegenerate inertia operator
${\mathcal I}: \mathfrak{g} \to \mathfrak{g}$ in the usual way:
\begin{gather*}
\forall \eta_1,\eta_2\in T_g G, \quad
(\eta_1,\eta_2)_g=\langle {\mathcal I}(\omega_1),\omega_2\rangle, \\
\mbox{where} \quad
\omega_1=g^{-1} \eta_1, \quad \omega_2=g^{-1} \eta_2 .
\end{gather*}
Let $y_1,\dots, y_n$ be independent left-invariant vector fields on $G$ generated by
some basis vectors $Y_1,\dots, Y_n$ in the algebra.
 Following \cite{VeVe1, VeVe2}, one can define
an \textit{LR system} on $G$ as a nonholonomic Lagrangian system $(G,l,D)$
where $l=\frac12(\dot g,\dot g)-v(g)$ is the Lagrangian with a left-invariant
kinetic energy and $D$ is
a {\it right-invariant\/} (generally nonintegrable) distribution on the tangent bundle $TG$.

The right-invariant distribution is determined by its restriction $\mathfrak{d}$ to
the Lie algebra as follows: $D_g=\mathfrak{d} \cdot g= g\cdot (g^{-1}\cdot
\mathfrak{d} \cdot g) \subset T_g G$, $\mathfrak{d}=$const.  Let
${\mathfrak h}=\mbox{span }( {\mathfrak h}_1,\dots, {\mathfrak h}_m)$ be the orthogonal
complement of $\mathfrak{d}$ with respect to $\langle \cdot,\cdot \rangle$ and
${\mathfrak h}_s=$const.
Then the right-invariant constraints can be written as
\begin{equation}
\omega\in g^{-1}\cdot \mathfrak{d} \cdot g, \quad {\mathrm{or}}\quad
f_s= \langle \omega,g^{-1}\cdot \mathfrak{h}_s\cdot g\rangle=0, \qquad
s=1,\dots, m, \label{dist}
\end{equation}
where $\omega=g^{-1}\cdot \dot g$.

The LR system $(G,l,D)$ can be described by the
Euler--Poincar\'e  equations
(also referred to as the Poincar\'e--Chetayev or Bolzano--Hamel equations) on the
product $\mathfrak{g}\times G$,
\begin{equation} \label{EPS}
\begin{aligned}
\frac {d}{d t} {\mathcal I}\omega
& = [{\mathcal I}\omega,\omega]
-y(v(g))+\sum^{m}_{s=1}\lambda_{s}\, g^{-1}\cdot \mathfrak{h}_s\cdot g\ , \\
\dot g &= g \omega,
\end{aligned}
\end{equation}
where $y(v)=(y_1(v),\dots, y_n(v))^T$ is the vector of Lie derivatives with
respect to the above left-invariant fields $y_1,\dots, y_n$,
and $\lambda_{s}$ are indefinite multipliers which can be found by differentiating (\ref{dist}).

These equations define a dynamical system on the whole
tangent bundle $TG$, and the right-invariant constraint functions $f_s$ in (\ref{dist}) are
its generic first integrals. Thus, the LR system $(G,l,D)$ itself can be regarded as
the restriction of the system (\ref{EPS}) onto $D\subset TG$.
(Also, the LR system with non-homogeneous right-invariant constraints $f_s=c_s\ne 0$
can be considered as a subsystem of (\ref{EPS}).)

For the case $v(g)=0$, system (\ref{EPS}) can be reduced to the form
\begin{equation} \label{EPS1}
\begin{aligned}
\frac {d}{d t} {\mathcal I}\omega
& = [{\mathcal I}\omega,\omega]+\sum^{m}_{s=1}\lambda_{s}{\mathcal F}_s, \\
\dot {\mathcal F}_s & = [{\mathcal F}_s,\omega],
\end{aligned}
\end{equation}
where
$
{\mathcal F}_s(g)=\partial f_s(\omega, g)/\partial \omega=g^{-1}\cdot \mathfrak{h}_s\cdot g.
$
This forms a closed system on the space
$(\omega, {\mathcal F}_1,\dots, {\mathcal F}_s)$.
\medskip

There is another way of describing LR systems, which is based
on the nonholonomic version of the Noether
theorem (see, e.g., \cite{AKN, FeKo2, BKMM}). Namely, as shown in \cite{VeVe2},
for $v(g)=0$, equations (\ref{EPS}) have the conservation
law $\frac{d}{dt} \pr_{\mathfrak{d}} (g\cdot {\mathcal I}\omega\cdot g^{-1})=0$,
which can be rewritten as
\begin{equation}
\frac{d}{dt} (\pr_{g^{-1}\cdot \mathfrak{d} \cdot g} {\mathcal I}\omega)=
[\pr_{g^{-1}\cdot \mathfrak{d} \cdot g} {\mathcal I}\omega,\omega].
\label{1.4}
\end{equation}
Next, for the case of non-homogeneous constraints $f_s=c_s$, one has \\
$\frac d{dt} ( \pr_{\mathfrak h} (g \omega g^{-1} ) ) =0$, which implies
$$
\frac d{dt} ( \pr_{g^{-1}{\mathfrak h}g} \omega )= [\pr_{g^{-1}{\mathfrak h}g}
\omega, \omega].
$$
Combining the above equations, we obtain the momentum equation
\begin{gather}
 \dot {\cal M}= [{\cal M},\omega],  \label{mom_eq} \\
{\cal M}=\pr_{g^{-1}\cdot \mathfrak{d}\cdot g} {\mathcal I}\omega+
\pr_{g^{-1} \cdot \mathfrak{h}\cdot g}\omega \, . \label{mom}
\end{gather}
As follows from (\ref{mom}),
the linear operator sending $\omega$ to $\cal M$ is nondegenerate,
and one can express  $\omega$ in terms of ${\mathcal M}$ and
the group coordinates $g$ uniquely. Thus (\ref{mom_eq}) together with
the kinematic equations $\dot g=g\omega$
represent a closed system of differential equations on the space $(\omega, g)$
or on $({\mathcal M}, g)$, which is equivalent to system (\ref{EPS}).
Since on $D\subset TG$ we have
$\pr_{g^{-1}\cdot \mathfrak{d}\cdot g} {\mathcal M}=\pr_{g^{-1}\cdot \mathfrak{d}\cdot g} {\mathcal I}\omega$, on this subvariety the system has the kinetic energy integral
$\frac 12 \langle {\mathcal M}, \omega\rangle
=\frac 12 \langle {\mathcal I}\omega, \omega\rangle$.
\medskip

Now let
$$
{\mathfrak d}=\mbox{span }({\mathfrak w}_1,\dots, {\mathfrak  w}_{n-m}), \quad
\langle {\mathfrak w}_k, {\mathfrak w}_s \rangle=\delta_{ks}
$$
and put $\mathcal W_k=g^{-1}\cdot{\mathfrak  w}_k\cdot g$. Then the above system
leads to a closed system of differential equations on the space $(\omega, \mathcal W_k)$
or on $({\mathcal M}, \mathcal W_k)$,
\begin{gather}
 \dot {\cal M} = [{\cal M},\omega], \quad \dot{\mathcal W}_k=[\mathcal W_k,\omega],
\label{mom_eq1} \\
{\cal M}= \omega+\sum_{k=1}^{n-m} \langle {\mathcal I}\omega -\omega, \mathcal W_k \rangle
\mathcal W_k. \nonumber
\end{gather}
The distribution $D$ is represented as an invariant subvariety of (\ref{mom_eq1})
given by the condition
$$
\omega -\sum_{k=1}^{n-m} \langle \omega, {\mathcal W}_k \rangle {\mathcal W}_k
\equiv
{\cal M}  -\sum_{k=1}^{n-m} \langle {\cal M} , {\mathcal W}_k \rangle {\mathcal W}_k =0.
$$

\paragraph{Reduction.}
Let the linear subspace $\mathfrak h$ be the Lie algebra of a subgroup $H\subset G$.
Furthermore, we suppose that the potential $v(g)$ is $H$--invariant.
Then the Lagrangian $l=\frac12(\dot g,\dot g)-v(g)$  and the
right-invariant distribution  $D$ are also invariant with respect to the left $H$--action.
(Notice that for $m>1$ the constraint functions $f_s$ themselves may not be $H$-invariant.)
In this case {\it the LR system $(G,l,D)$ can naturally be regarded
as a generalized Chaplygin system.}

Consider the homogeneous space $Q=H \backslash G$ of left cosets
$\{Hg\}$. The distribution $D$ can be regarded as a principal connection of the principal bundle:
\begin{equation*} \begin{array}{cccc}
H & \longrightarrow & G &  \\
&  & \downarrow & \pi \\
&  & Q= H \backslash G &
\end{array}.
\end{equation*}

The Lagrange-d'Alembert equation (\ref{1.1a})
is $H$--invariant and it reduces to a second order equation on $Q$.
In order to write the reduced equations in a simple form,
we identify $\mathfrak{g}$ and $\mathfrak{g}^*$ by the $Ad_G$-invariant scalar
product $\langle \cdot,\cdot \rangle$, and the spaces $TQ$, $T^*Q$ by
the {\it normal\/} metric induced by the bi-invariant metric on $G$. Next,
consider the momentum maps
\begin{equation*}
\phi: TG \cong T^*G\to \mathfrak{g}, \quad \Phi: TQ\cong T^*Q \to
\mathfrak{g}
\end{equation*}
of the natural \textit{right} actions of $G$ on $T^*G$ and $T^*Q$, respectively.
We have
$$
\phi(X)=g^{-1}\cdot X, \quad X\in T_g G
$$
and the map $\Phi$ can be considered as a restriction of $\phi$ to $D$.

The {\it reduced Lagrangian} is by definition the constrained Lagrangian
$$
l_c(g,\dot g)=\frac12\langle \pr_{g^{-1}\mathfrak d g}
\mathcal I(\phi(g,\dot g)),\phi(g,\dot g)\rangle-v(q)
$$
considered on the orbit space $H\backslash D=T(H\backslash G)$.
It follows that the reduced Lagrangian is simply given by
\begin{equation*}
L(q,\dot q)=\frac12 \langle {\mathcal I} \Phi(q,\dot q), \Phi(q,\dot q) \rangle-V(q),
\end{equation*}
where $q=\pi(g)$ are local coordinates on $Q$ (which may be
redundant) and $V(q)=v(g)$. For $v=V=0$, this Lagrangian
describes a metric which we shall denote by $ds^2_{{\mathcal I},D}$.

The reduced system on $T Q$ is defined by the following proposition, which
appears to be a special case
of the general nonholonomic reduction procedure described in \cite{Koi, BKMM}.

\begin{prop} \label{pro}
The reduced Lagrange--d'Alembert equation describing the motion of the LR system
$(G,l,D)$ takes the form
\begin{equation}
\left( \frac{\partial L}{\partial q} - \frac{d}{dt}
\frac{\partial L}{\partial \dot q}, \xi\right) =
\langle {\mathcal I}\Phi(q,\dot q), \pr_{g^{-1}\mathfrak{h} g}
 [\Phi(q,\dot q),\Phi(q, \xi)] \rangle, \quad \textup{for all }
\xi\in T_q Q,   \label{1.2}
\end{equation}
where $\pr_{g^{-1}\mathfrak{h} g}: \mathfrak{g}\to g^{-1}\mathfrak{h} g$
is the orthogonal projection and $q=\pi(g)$.
\end{prop}

As a result, (\ref{1.2}) leads to a system of Lagrange equations on $TQ$ with some
extra terms. Note that this system always has the energy integral
\begin{equation*}
E(q,\dot q)=\frac12 \langle {\mathcal I}\Phi(q,\dot q),\Phi(q,\dot q)\rangle +V(q).
\end{equation*}
\medskip

\noindent{\it Proof of Proposition} \ref{pro}.
First, we need to describe the curvature of the principal connection associated to the
distribution $D$.
Let $X_1,X_2\in T_g G$. Then the horizontal and vertical components of $X_i$ have the form
$$
X_i^h=g\cdot\pr_{g^{-1}{\D}g} \phi(X_i), \quad
X_i^v=g\cdot\pr_{g^{-1}{\h}g} \phi(X_i).
$$
Next, if $\bar X_1, \bar X_2$ are right invariant extensions
of $X_1$ and $X_2$, then $[\bar X_1,\bar X_2]_g=-g\cdot [\phi(X_1),\phi(X_2)]$.
(Here the first square brackets denote the commutator of vector fields,
and the second ones represent the commutator in the algebra $\g$.) Thus, the curvature is
$$
B(X_1,X_2)=-[\bar {X_1}^h,\bar {X_2}^h]_g^v=g\cdot \pr_{g^{-1}\h g}
[\pr_{g^{-1}\D g}\phi(X_1),\pr_{g^{-1}\D g}\phi(X_2)].
$$
Therefore the right hand side of (\ref{1.1a}) is equal to
$$
\left(\frac{\partial l}{\partial \dot g},g\cdot \pr_{g^{-1}\h g}[\omega,\phi(\eta)]\right)=
\langle {\mathcal I}\omega,\pr_{g^{-1}\h g}[\omega,\phi(\eta)]\rangle, \quad
\omega=g^{-1}\cdot \dot g=\phi(\dot g).
$$
Combining the above expressions, we arrive at the right hand side of (\ref{1.2}).
\medskip

\paragraph{Reduced momentum equation.}
Similar to the original LR systems,
in the absence of potential forces one can describe reduced LR systems
on $T^*Q$  in terms of a momentum equation as well.

Namely, let us now identify $\{p\}=T^*_q Q$ and $\{\dot q\}\in T_q Q$ by the metric
$ds^2_{\mathcal I, D}$, i.e., we put $p=\partial L(q,\dot q)/\partial \dot q$, and
also identify  the spaces  ${\mathfrak g}=\{\omega\}$ and
${\mathfrak g}^*=\{ {\mathcal M}\}$ via relation (\ref{mom}).

Next, introduce the momentum map $\Phi^*\, : T^*Q \to \g^*$, $(q,p) \to {\mathcal M}$
by setting
$$
\Phi^*(q,p) = \Phi(q,\partial L(q,\dot q)/\partial \dot q),
$$
where $\partial L(q,\dot q)/\partial \dot q$ is considered as an
element of $T_q Q$ (via identification given by the normal metric).

The map is correctly defined because
\begin{equation} \label{*}
\Phi^*(q,p)|_{p=\partial L(q,\dot q)/\partial \dot q}
= \pr_{g^{-1}\mathfrak d g} {\mathcal I} \Phi(q,\dot q)
= \pr_{g^{-1}\mathfrak d g} {\cal M} .
\end{equation}
Indeed, a preimage of
$\partial L(q,\dot q)/\partial \dot q$ in $D_g\subset T_g G$, $g\in \pi^{-1}(q)$
can be chosen in form $\partial l_c(g,\dot g)/\partial \dot g$,
$\pi_*(\dot g)=\dot q$. Therefore, we have
\begin{gather*}
\Phi(q,\partial L(q,\dot q)/\partial \dot q)
=\pr_{g^{-1} \mathfrak d g} \phi(g,\partial l_c(g,\dot g)/\partial \dot g)\\
= \pr_{g^{-1}\mathfrak d g}\left(
 g^{-1}\cdot \frac{\partial l_c(g,\dot g)}{\partial \dot g}\right)
= \pr_{g^{-1}\mathfrak d g} g^{-1} (g\mathcal I g^{-1}) \dot g
= \pr_{g^{-1}\mathfrak d g} {\mathcal I}\omega  ,
\end{gather*}
which establishes the first equality in (\ref{*}).
The second equality follows from (\ref{mom}).
\medskip

Since  the linear subspace $g^{-1}\mathfrak d g \subset \mathfrak
g$ is $H$-invariant, it depends only on $q\in Q$. Thus, the system
(\ref{mom_eq1})  represented in terms of $\omega$ can be regarded
as a flow on the quotient manifold $H\backslash TG\cong Q \times
{\mathfrak g}$ obtained from $TG \cong G\times {\mathfrak g}$ by
factorization by $H$. The same system represented in terms of
${\mathcal M}$ leads to a system on $Q \times {\mathfrak g}^*$.

Relations between the above manifolds are described by the commutative diagram below,
where the vertical arrows denote the corresponding inclusions and
$\tilde \Phi, \tilde\Phi^*$ are the extensions
of the momentum maps $\Phi, \Phi^*$ respectively.
$$
\begin{CD}
TG\, \cong  \, G \times {\mathfrak g}  @ >  H\backslash >>
Q \times {\mathfrak g}  @ > \mbox{(2.8)} >>  Q \times {\mathfrak g}^*     \\
 @  A   f_s(g,\omega)=0  AA     @  A   \tilde {\Phi } AA @  A \tilde{\Phi}^* AA \\
D\, \cong \, G\times {\mathfrak d}  @ > H \backslash   >>  T Q @  >
p=\partial L(q,\dot q)/\partial \dot q >> T^*Q
\end{CD}
$$
For a fixed $q$, the map $\Phi^*$ establishes a bijection between
the subspace $g^{-1}\mathfrak d g \subset \mathfrak g^*$  and the
cotangent space $T^*_q Q$.

Now, applying (\ref{1.4}) and (\ref{*}), we arrive at the reduced momentum equation
\begin{equation}
\frac{d}{dt}\Phi^*(q,p)=[\Phi^*(q,p),\Phi(q,\dot q)],
\label{**}
\end{equation}
where $\dot q=\dot q(q,p)$ is determined from
$p=\partial L(q,\dot q)/\partial \dot q$. This leads to a system of equations
on $T^* Q$, which are equivalent to the Lagrange equations on $TQ$ obtained from
(\ref{1.2}).

As a consequence of the momentum equation (\ref{**}), we also obtain the following
result.
\begin{prop}
In the absence of potential forces the reduced LR system on $T^*Q$
always has a set of first integrals
$\mathcal A=\{f\circ \Phi^*, \, f\in \mathbb R[\mathfrak g]^G\}$,
where $\mathbb R[\mathfrak g]^G$ is the algebra of $Ad_G$ invariants on $\mathfrak g$.
\end{prop}

The number of independent functions in $\mathcal A$ is equal to
the number of independent $G$--invariant functions on $T^*Q$,
that is to $\dim \mathrm{pr}_\mathfrak d(\mathrm{ann}(\xi))$, for a generic $\xi\in \mathfrak d$
(see \cite{BJ}).
Here $\mathrm{ann}(\xi)=\{\eta\in\mathfrak g,\; [\xi,\eta]=0\}$.
If $Q=H \backslash G$ is a symmetric space, this number
is equal to the rank of $Q$.
\medskip


\section{Invariant measure and time rescaling}
One of the remarkable properties of LR systems is the existence of an invariant
measure, which puts them rather close to Hamiltonian systems.

\begin{thm} \label{IM1} \textup{(\cite{VeVe1, VeVe2}).}
The LR system (\ref{EPS1}) on the space
$(\omega, {\mathcal F}_1,\dots, {\mathcal F}_s)$
possesses an invariant measure with density
\begin{equation} \label{measure1}
\mu= \sqrt{ \det ( {\mathcal I}^{-1} |_{g^{-1} \mathfrak{h} g} ) }
\equiv \sqrt{ \det \langle{\mathcal F_s},{\mathcal I}^{-1} {\mathcal F_l}\rangle }, \qquad
s,l=1,\dots, m,
\end{equation}
where ${\mathcal I}^{-1}|_{g^{-1} \mathfrak{h} g}$
is the restriction of the inverse inertia tensor to the linear space
$g^{-1}\mathfrak{h} g \subset {\mathfrak g}$.
\end{thm}
\medskip

The alternative description of LR systems given by the momentum equation
(\ref{mom_eq}) leads to another expression for invariant measure.
\begin{thm} \label{IM2}
The LR system defined by the momentum equation (\ref{mom_eq1})
has the invariant measure
\begin{gather} \label{measure2}
\tilde\mu \, d\omega\wedge d \mathcal W_1 \wedge \cdots
\wedge d \mathcal W_{n-m}= \tilde\mu^{-1} \,
d {\mathcal M}\wedge d\mathcal W_1 \wedge \cdots \wedge d\mathcal W_{n-m}\, , \\
\tilde\mu= \bigg | \frac{\partial {\mathcal M}}{\partial\omega} \bigg |^{1/2}=
\sqrt{ \det ( {\mathcal I}|_{g^{-1} \mathfrak{d} g} ) }
\equiv \sqrt{ \det \langle \mathcal W_i,{\mathcal I} \mathcal W_j \rangle }, \label{tilded_den} \\
i,j=1,\dots, n-m,\nonumber
\end{gather}
where ${\mathcal I}|_{g^{-1} \mathfrak{d} g}$
is now the restriction of the inertia tensor to the linear space
$g^{-1}\mathfrak{d} g \subset {\mathfrak g}$.
\end{thm}
Expressions (\ref{measure1}) and (\ref{tilded_den}) involve complimentary basis
vectors in $g^{-1} \mathfrak{g} g$. In this sense
the densities $\mu$ and $\tilde\mu$ given by the above theorems are {\it dual}.
\medskip

\noindent{\it Proof of Theorem \ref{IM2}}.
First note that the systems (\ref{EPS}) and (\ref{mom})
can be extended to one and the same system on the space
$(\omega,{\mathcal F}_1,\dots, {\mathcal F}_m, \mathcal W_1, \dots, \mathcal W_{n-m})$ by
adding evolution equations for $\mathcal W_j$ and ${\mathcal F}_i$ respectively.
The resulting system has an invariant measure whose density can differ from those
of the original systems only by constant factors.
Hence the functions $\mu$ in (\ref{measure1}) and  $\tilde\mu$ in (\ref{measure2})
can be different only by a constant multiplier.

Next, note that in an appropriate $g$-dependent orthogonal basis in the algebra $\mathfrak g$,
the Jacobian matrix $\partial {\mathcal M}/\partial\omega$ has the following block
structure
$$
\frac{\partial {\mathcal M}}{\partial\omega}
=\begin{pmatrix} {\bf I}_{n-m} & 0 \\
    0 & 0 \end{pmatrix} {\mathcal I} +
 \begin{pmatrix} 0 & 0 \\
    0 & {\bf I}_{m} \end{pmatrix} \equiv
\begin{pmatrix} {\mathcal I}|_{g^{-1} \mathfrak{d} g}  & {\cal S} \\
    0 & {\bf I}_{m}  \end{pmatrix} ,
$$
where ${\bf I}_{n-m},{\bf I}_{m}$ are unit matrices of dimension $(n-m)\times (n-m)$
and $m\times m$ respectively, and $\cal S$ is some $(n-m)\times m$-matrix.
In the same basis one has
$$
\frac{\partial {\mathcal M}}{\partial\omega} {\mathcal I}^{-1}
=\begin{pmatrix} {\bf I}_{n-m} & 0 \\
    0 & 0 \end{pmatrix} +
 \begin{pmatrix} 0 & 0 \\
    0 & {\bf I}_{m} \end{pmatrix} {\mathcal I}^{-1}  \equiv
\begin{pmatrix} {\bf I}_{n-m}  & 0 \\
    {\cal U}  &  {\mathcal I}^{-1}|_{g^{-1} \mathfrak{h} g} \end{pmatrix} ,
$$
with some $m\times (n-m)$-matrix $\cal U$.
Comparing the right hand sides of these two expressions with (\ref{measure1}),
we obtain the following chain:
\begin{equation} \label{duality}
\mu^2= \det ( {\mathcal I}^{-1} |_{g^{-1} \mathfrak{h} g} ) =
 \left | \frac{\partial {\mathcal M}}{\partial\omega} {\mathcal I}^{-1} \right|
=\det ({\mathcal I}^{-1}) \left | \frac{\partial {\mathcal M}}{\partial\omega} \right| =
\det ({\mathcal I}^{-1}) \det ( {\mathcal I} |_{ g^{-1} {\mathfrak d} g} ) .
\end{equation}
Hence, we can choose the density $\tilde\mu$ in the form (\ref{tilded_den}).

Finally, taking into account the relation
$$
d\omega\wedge d\mathcal W_1 \wedge \cdots \wedge d\mathcal W_{n-m}
=\left | \frac{\partial {\mathcal M}}{\partial\omega} \right| ^{-1}
 d {\mathcal M}\wedge d\mathcal W_1 \wedge \cdots \wedge d\mathcal W_{n-m}
$$
and using (\ref{tilded_den}),
we come to the equality in (\ref{measure2}). The theorem is proved.
\medskip

As shown in \cite{VeVe2}, Theorem \ref{IM1} implies that the original nonholonomic
system (\ref{EPS1}) on the left trivialization ${\mathfrak g}\times G$ of $TG$ has
the invariant measure $\mu(g)\, d\omega \wedge dg$.

\paragraph{Reduced invariant measure.} Now we proceed to reduced LR systems.
As a natural consequence of the above theorems, we have

\begin{thm} \label{red_measure}
The reduced LR system (\ref{1.2})
(or, after the Legendre transformation, the system (\ref{2.1})
on $T^*(H\backslash G)$) possesses an invariant measure.
\end{thm}

Note that the reduction of a generic Chaplygin system may not have this property
(see \cite{CCLM}).

The proof of Theorem \ref{red_measure} consists of two steps. First, it is seen that
the restriction of the LR system (\ref{EPS1}) onto the distribution $D\subset TG$ has an invariant measure. Indeed, the volume form on the tangent bundle admits the decomposition
\begin{equation}\label{decomp}
d\omega \wedge dg = \theta (g) \, df_1 \wedge\cdots\wedge df_m \wedge \varPi\, ,
\end{equation}
where $f_s(\dot g,g)$ are the constraint functions in (\ref{dist}), $\theta (g)$ is a function,
and $\varPi$ is a volume form on $D$. Since the 1-forms $df_s$ are independent on $TG$,
$\theta (g)$ does not vanish on $G$.

Let  ${\mathcal L}_*$ be the Lie derivative with respect to the nonholonomic flow
(\ref{EPS1}).
Since the functions $f_s(\dot g,g)$ are its generic first integrals,  we have
${\mathcal L}_* d\,f_s = d (\dot f_s)=0$, $s=1,\dots,m$.  As a result, from the
condition ${\mathcal L}_* (\mu \, d\omega \wedge dg)=0$ and (\ref{decomp})
we obtain
$df_1 \wedge\cdots\wedge df_m \,{\mathcal L}_* (\mu \theta\, \varPi)=0$.
Hence, the restriction of the flow onto $D$ has the invariant measure
$\mu(g) \theta(g) \, \varPi$.

Notice that one can always choose $\varPi$ to be $H$-invariant. In this case, since
the form $d\omega \wedge dg$ is $G$-invariant, whereas
the wedge product $df_1 \wedge\cdots\wedge df_m$ and $\mu(g)$ are $H$-invariant,
the density $\mu(g) \theta(g)$ of the restricted measure is also $H$-invariant and goes
down to $Q$.

The second step is based on the following general lemma.
(Although it is quite natural, we could not find it in the literature.)

\begin{lem} \label{red*} Suppose a compact group $\mathfrak G$
acts freely on a manifold $N$ with local coordinates z, and
there is a $\mathfrak G$--invariant dynamical system $\dot z=Z(z)$ on $N$.
If this system has an invariant measure (which is not necessarily $\mathfrak G$-invariant), then the reduced system on the quotient manifold $N/\mathfrak G$ also has an invariant measure.
\end{lem}

Now, identifying the group $\mathfrak G$ and the manifold $N$ with $H$ and $D$ respectively, we arrive at Theorem \ref{red_measure}.
\medskip

\noindent{\it Proof of Lemma\/} \ref{red*}  The manifold $N$ can be locally represented
as a direct product $\mathbb R^k\{x\}\times \mathfrak G$, where
$x$ is a local coordinate system on $N/\mathfrak G$, so that
the $\mathfrak G$--action and the dynamical system take the form
\begin{gather*}
a\cdot (x,g)=(x,ag),\quad a\in \mathfrak G, \\
\mbox{and} \quad \dot x=X(x), \quad \dot g=Y=g\cdot \xi(x),
\quad \xi(x)\in {\mathfrak g}=T_{Id}\mathfrak G,
\end{gather*}
respectively.

Let $\Theta$ be an invariant measure of the original system on $N$,
$\mu$ be a bi-invariant volume form on $\mathfrak G$,
$\sigma$ be a volume form on $N/\mathfrak G$ and $\sigma_x$
be its local representation in $x$--coordinates.
Then the invariant measure on $N$ locally has the form
$\Theta =f(x,g)\, \mu\wedge\sigma_x$. Thus
\begin{equation}
{\mathcal L}_Z f(x,g)\, \mu\wedge\sigma_x=
Z(f)\,\mu\wedge\sigma_x+f\cdot ({\mathcal L}_Z \mu)\wedge \sigma_x+
f\mu\wedge({\mathcal L}_Z\sigma_x)=0,
\label{r2}
\end{equation}
where ${\mathcal L}_Z$ is the Lie derivative with respect to the
flow $Z$. Since $d\sigma_x=d_x\sigma_x=0$ and $d\mu=d_g\mu=0$, we
have
\begin{equation}
{\mathcal L}_Z\sigma_x=(d\circ i_Z)\sigma_x+(i_Z\circ d)\sigma_x=
d(i_Z\sigma_x)=d_x(i_X\sigma_x)={\mathcal L}_X\sigma_x,
\label{r3}
\end{equation}
\begin{equation}
{\mathcal L}_Z\mu=(d\circ i_Z)\mu+(i_Z\circ d)\mu=d(i_Z\mu)=d(i_Y\mu)=(d_x+d_g)(i_Y\mu).
\label{r4}
\end{equation}
For a fixed $x$, $Y=Y(x)$ is a {\it left-invariant} vector field on
$\mathfrak G$,  whereas the corresponding flow on
$\mathfrak G$ is {\it right-invariant}. Since $\mu$ is
bi-invariant, we have
${\mathcal L}_Y\mu=d_g(i_Y\mu)=0.$
Also, it is obvious that $d_x(i_Y\mu)\wedge\sigma=0$.
Therefore, taking into account (\ref{r2}--\ref{r4}), we get
\begin{equation}
Z(f)\mu\wedge\sigma_x+f\mu\wedge({\mathcal L}_X\sigma_x)=0
\label{r6}
\end{equation}

Now we introduce the ``averaged'' density $\bar f(x)=\int_{\mathfrak G} f(x,g)\mu$,
which, as we shall see below, has the following property
\begin{equation}
\int_{\mathfrak G} Z(f)\mu=X\left(\int_{\mathfrak G} f\mu\right)=X(\bar f).
\label{r7}
\end{equation}
Then, by integration of (\ref{r6}), we obtain
$
X(\bar f)\sigma_x+\bar f {\mathcal L}_X\sigma_x=0.
$
As a result, the reduced system preserves the volume form $\bar f(x)\sigma_x$.

We stress that the above procedure does not depend
on the choice of the local coordinates on $N/\mathfrak G$. Indeed,
let $y=y(x)$ be another coordinate system. Then
$$
\Theta=h(y,g)\, \mu\wedge\sigma_y=
h(y(x),g)\mu\wedge \det\left(\frac{\partial y}{\partial x}\right)\sigma_x=
f(x,g)\, \mu\wedge\sigma_x,
$$
and after integration we have
$\bar f(x)\sigma_x=\bar h(y) \sigma_y$.

It remains to prove (\ref{r7}).
We have $Z(f)=X(f)+Y(f)$ and
$
\int_{\mathfrak G} X(f) \mu=
X\left(\int_{\mathfrak G} f\mu\right).
$
Therefore the relation (\ref{r7}) is equivalent to
\begin{equation}
\int_{\mathfrak G} Y(f)\mu=0.
\label{r8}
\end{equation}
To check the latter relation, we fix $x$. Then ${\mathcal L}_Y (f\mu)=Y(f)\mu+f{\mathcal L}_Y\mu$
and, on the other hand, ${\mathcal L}_Y (f\mu)=d_g(i_Y(f\mu))$.
Since ${\mathcal L}_Y\mu=0$, we get
$Y(f)\mu=d_g(i_Y(f\mu))$, and (\ref{r8}) follows from the Stokes theorem.
The lemma is proved.

\paragraph{Chaplygin reducing multiplier.} Here we continue with the reduced LR systems.
However, all considerations
hold for an arbitrary generalized Chaplygin system with the
Lagrangian of the natural mechanical type.
Let $q_1,\dots,q_k$ be some local coordinates on
the homogeneous space  $Q$ and $p_1,\dots,p_k$,
$p_i=\partial L/\partial \dot q_i$ be canonically conjugated momenta which provide
coordinates on the cotangent bundle $T^*Q$. Let $g_{ij}$ denote metric tensor of
$ds^2_{{\mathcal I},D}$ and $g^{ij}$ give the dual metric on $T^*Q$.

The reduced Lagrangian is $L(q,\dot q)=\frac12 \sum g_{ij} \dot q_i\dot q_j -V(q)$.
We also introduce the Hamiltonian function $H(q,p)=\frac12 \sum g^{ij} p_ip_j+V(q)$
(the usual Legendre transformation of $L$).
Then (\ref{1.2}) can be rewritten as a first-order dynamical system on $T^*Q$:
\begin{equation}
\dot q_i=\frac{\partial H}{\partial p_i},\quad
\dot p_i=-\frac{\partial H}{\partial q_i}+\Pi_i(q,p), \qquad i=1,\dots,k.  \label{2.1}
\end{equation}
 The functions $\Pi_i$ are quadratic in momenta and can be regarded as
non-Hamiltonian perturbations of the equations of motion of a particle on $Q$.

Now consider time substitution $d\tau={\mathcal N} (q)dt$, where
${\mathcal N}(q)$ is a differentiable nonvanishing function on $Q$,
and denote $q^{\prime}={dq}/{d\tau}$.
Then we have the following commutative diagram:
\begin{equation*}
\begin{CD}
 TQ\{q,\dot q\} @ > q^{\prime}=\dot q/{\mathcal N}(q) >> TQ\{q,q^{\prime}\} \\
 @ V p=g \dot q VV   @ VV \tilde p={\mathcal N}^2 g q^{\prime} V \\
 T^*Q\{q,p\}  @ >  \tilde p=\mathcal{\ N} p >>  T^*Q\{q,\tilde p\} .
\end{CD}
\end{equation*}

The Lagrangian and Hamiltonian functions in the coordinates
$\{q,q^{\prime}\} $ and $\{q,\tilde p\}$ take the form
\begin{equation*}
L^*(q,q^{\prime})= \frac{1}{2}\sum \mathcal{\ N}^2 g_{ij}
q_i^{\prime}q_j^{\prime}-V(q), \quad H^*(q,\tilde p)=\frac12\sum \frac{1}
{\mathcal{\ N}^2} g^{ij} \tilde p_i \tilde p_j+V(q).
\end{equation*}

There is a remarkable relation between the existence of an invariant measure
of the reduced system (\ref{2.1}) and its reducibility to a Hamiltonian form.

\begin{thm} \label{Th_reduce} \begin{description}
\item{1).}
Suppose that after the time substitution $d\tau=\N(q) dt$
the equations (\ref{2.1}) become Hamiltonian,
\begin{equation}
q_i'=\frac{\partial H^*}{\partial \tilde p_i}, \quad
\tilde p_i'= -\frac{\partial H^*}{\partial q_i}\, .
\label{x2}
\end{equation}
Then the original system (\ref{2.1})  has the invariant measure
$$
\N(q)^{k-1}\, dp_1\wedge \cdots \wedge dp_k\,\wedge  dq_1 \wedge\cdots\wedge dq_k
\equiv \N(q)^{k-1}\,\varOmega^k,
$$
where $\varOmega=\sum dp_i\wedge dq_i$ is the standard symplectic form on $T^*Q$.

\item{2).} For $k=2$, the above statement can also be inverted:
the existence of the invariant measure with the density
$\N(q)$ implies that in the new time $d\tau=\N(q)dt$,
the system (\ref{2.1})  gets the Hamiltonian form (\ref{x2}).
\end{description}
\end{thm}

In nonholonomic mechanics the factor $\N$ is known as the {\it reducing multiplier\/},
item 2) of this theorem is referred to  as {\it Chaplygin's reducibility theorem\/}
(see \cite{Ch1, Ch2, Ch} or section III.12 in \cite{NeFu}).
Item 1) of the theorem was implicitly formulated in \cite{St, CCLM}.

\begin{remark} {\rm Notice that for $k>2$, the multiplier $\N (q)$ and
the density of the invariant measure of system (\ref{2.1})  do not
coincide.}\end{remark}

On the other hand, the paper \cite{BMK} gives examples of reducibility of
nonholonomic systems to Hamiltonian ones with respect to {\it nonlinear\/} Poisson brackets.
In some of these examples, even for $k=2$ the reducing factor and the
density of invariant measure are different.
\medskip

\noindent{\it Proof of item 1) of Theorem\/} \ref{Th_reduce}.
For simplicity we shall use the vector notation $p=(p_1,\dots,p_k)$, $q=(q^1,\dots,q^k)$, etc.
Let $G$ be the matrix $(g^{ij})$. Then $\dot q=Gp$, $H=\frac12(Gp,p)$,
$H^*=\frac{\N^2}{2}(G\tilde p,\tilde p)$.

The equations (\ref{x2})
in the original time $t$ take the form
\begin{equation}
\dot q= \N \nabla_{\tilde p} H^*(q,\tilde p),\quad
\dot {\tilde p}= - \N\nabla_q H^*(q,\tilde p).
\label{x1}
\end{equation}

Equations (\ref{2.1}) have an invariant measure with density $f$ if
\begin{equation}
(\nabla_q, f\nabla_p H)+(\nabla_p,f(-\nabla_q H+\Pi))=0.
\label{x7}
\end{equation}

For $f$ which depend only on $q$--coordinates, we have
$
(\nabla_p,\Pi)+(\nabla_q \ln f,G p)=0,
$
or equivalently
\begin{equation}
d(\ln f)+\alpha=d(\ln f)+(A,dq)=0,
\label{x3}
\end{equation}
where $(\nabla_p,\Pi)=(A,\dot q)=\alpha(\dot q)$.
In particular, the one-form $\alpha$ is closed.

We shall prove that the function $f(q)=\N^{k-1}(q)$ satisfies equations (\ref{x3}).
Since $\tilde p=\N p$ we have
$\N\dot p+\dot\N p=\dot{\tilde p}$.
Therefore, using equations (\ref{x1}) we obtain
\begin{equation}
\dot p=\N^{-1}\dot{\tilde p}-\dot\N(q) p=-\nabla_q H^*(q,\tilde p)-(\nabla_q\N,Gp) p.
\label{x4}
\end{equation}
Also, one can easily see that
$
\nabla_q H^*(q,\tilde p)=\nabla_q H(q,p)-\N^{-1}(Gp,p)\nabla_q\N.
$
Thus, comparing (\ref{2.1}) and (\ref{x4}) we get
\begin{equation}
\Pi(q,p)=\N^{-1}(Gp,p)\nabla_q\N-\N^{-1}(\nabla N,Gp) p
\label{x5}
\end{equation}

Using (\ref{x5}) we see
$$
(\nabla_p,\Pi)=\frac{1}{\N}\left(2(\nabla_q\N, Gp)-k(\nabla_q\N,Gp)-(\nabla_q\N,Gp)\right)
=\frac{1-k}{\N}(\nabla_q\N, Gp).
$$
Hence $\alpha=-d\ln(\N^{k-1})$.

\medskip
As mentioned above, item 2) of the theorem is just a reformulation of Chaplyginïs
reducibility theorem in \cite{Ch2}.
\medskip

Clearly, the density of an invariant measure of a generic dynamical system depends on the
choice of local coordinates on the phase space. However, in the case of a system on
the cotangent bundle $T^* Q$ the density is invariant with respect to changes of
coordinates on $Q$, since the symplectic form $\varOmega$ and the measure itself are invariant
with respect to contact transformations.

\begin{remark} {\rm The paper \cite{St} (see also \cite{CCLM}) contains
a nontrivial observation about the density of the
invariant measure, which in our terms reads as follows.
Let a function $f(q,p)$ be a solution of (\ref{x7}) in case of the absence of
the potential ($V(q)=0$). Then one can check that
the function $f_0(q)=f(q,0)$ also satisfies condition
(\ref{x7}), i.e., it is a solution of (\ref{x3}).
In other words, if the reduced system (\ref{2.1})
has an invariant measure for $V=0$, one can take this measure
to be of the form $f(q) \varOmega^k$.
Then, since (\ref{x3}) does not depend on the potential,
the reduced system (\ref{2.1}) has the same invariant measure
in the presence of the potential field $V(q)$ as well.}
\end{remark}

\section{Veselova system on $T\, SO(3)$,
the Neumann system and a geodesic flow on $S^2$}

The most descriptive illustration of an LR system is the
\textit{Veselova problem} on the  motion of a rigid body
about a fixed point under the action of the nonholonomic constraint
\begin{equation}
\label{classical}
(\Omega,\gamma )=0 .
\end{equation}
Here $\Omega$ is the vector of the angular velocity in the body frame,
$\gamma$ is a unit vector which is fixed
in a space frame, and $(\, ,\, )$ denotes the scalar product in ${\mathbb R}^3$ \cite{VeVe1}.
Geometrically this means that the projection of the angular velocity of the body to
a fixed vector must zero.

This setting should not be confused with the nonholonomic {\it Suslov problem\/},
when the analogous constraint is defined by a vector fixed {\it in the body frame\/}
(\cite{Bl_Z, FeKo2, Jo2}).

The equations of motion in the moving frame in the presence of a potential field
$V=V(\gamma)$ have the form
\begin{eqnarray}
{\mathcal I}\dot \Omega &=& {\mathcal I}\Omega \times \Omega + \gamma\times\frac{\partial V}
{\partial\gamma}+\lambda \gamma,  \notag \\
\dot \gamma &=& \gamma \times \Omega,
\label{3.1}
\end{eqnarray}
where $\mathcal I$ is the inertia tensor of the rigid body, $\times$ denotes the vector
product in ${\mathbb R}^3$, and  $\lambda$ is a Lagrange
multiplier chosen such that $\Omega(t)$ satisfies the constraint (\ref{classical}),
\begin{equation} \label{la}
\lambda= - \frac {({\mathcal I}\Omega \times \Omega + \gamma\times {\partial V}/
{\partial\gamma}, {\mathcal I}^{-1}\gamma)}{({\mathcal I}^{-1}\gamma,\gamma ) }\, .
\end{equation}

The Veselova system (\ref{classical}), (\ref{3.1}) is an LR system
on the Lie group $SO(3)$, which is the configuration space of the
rigid body motion. After identification of the Lie algebras
$(\mathbb{R}^3,\times)$ and $(so(3),[\cdot,\cdot])$, the operator
$\mathcal I$ induces the left-invariant metric $ds^2_I$. The
angular velocity vector $\Omega$ correspond to $g^{-1}\dot g$, the velocity
in the left trivialization $TSO(3)\cong SO(3)\times so(3)$, and
the Lagrangian function equals $\frac 12 (\Omega, {\mathcal
I}\Omega)-V(\gamma)$.  The fixed vector in the space corresponds
to the right-invariant vector field $\gamma_g=g\cdot (g^{-1}\cdot
\mathfrak h \cdot g)\in T_g SO(3)$, $\mathfrak h\in so(3)$, and
the nonholonomic constraint (\ref{classical}) has the form
$\langle g^{-1} \cdot \mathfrak h \cdot g, \Omega\rangle=0$.

Equations (\ref{3.1}), (\ref{la}) also define a dynamical system on the
space $\{\Omega,\gamma\}=so(3)\times {\mathbb R}^3$, and the constraint function
$(\Omega,\gamma)$ appears as its first integral.
As noticed in \cite{VeVe1},  this system has an invariant measure with density
$\sqrt{ ({\mathcal I}^{-1}\gamma,\gamma )}$. Apart from the above constraint,
it always has the geometric integral $(\gamma,\gamma)$.
According to \cite{Fed_nonholonomic}, for $V(\gamma)=0$ there are
two other independent integrals
\begin{equation}\label{ints}
\frac 12 (\Omega, {\mathcal I}\Omega)- (\Omega,\gamma) ({\mathcal I}\Omega,\gamma), \quad
\frac 12 ({\mathcal I}\Omega- ({\mathcal I} \Omega, \gamma)\gamma + (\Omega,\gamma)\gamma )^2,
\end{equation}
the first expression being an analog of the so called {\it Jacobi--Painlev\'e\/} integral
which replaces the energy integral in some systems with nonstationary constraints.

On the constraint subvariety (\ref{classical}), these functions
reduce to the energy integral $F_1=\frac12 ({\mathcal I}\Omega,\Omega)$ and an
additional integral
$$
F_2=\frac12 (I\Omega, I\Omega)- \frac12 (I\Omega,\gamma)^2 ,
$$
which was found in \cite{VeVe1}.

As a result, by the \textit{Euler--Jacobi theorem} (see e.g., \cite{AKN}), the above system
is solvable by quadratures on the whole space $so(3)\times {\mathbb R}^3$.
Note that analogous integrable LR systems on the group $SL(2,{\mathbb R})$ and the
Heisenberg group are studied in \cite{Jo1}.

As shown in \cite{VeVe2}, in the case of the absence of the potential
the Veselova system (\ref{3.1}), (\ref{classical})
can be explicitly integrated by relating it to the classical Neumann system.

\begin{thm}\label{LR->N3} \textup{(\cite{VeVe2})}. Let  $\gamma(t)$ be a solution of
equations (\ref{3.1}), (\ref{classical}) with $V(\gamma)=0$ and with energy constant $F_1=h$.
Then after time reparameterization
\begin{equation*}
d\tau_1= \sqrt{\frac{2 h \det {\mathcal I}^{-1} } {( {\mathcal I}^{-1}\gamma,\gamma)} }\, dt
\end{equation*}
the unit vector $q=\gamma$ is a solution of the Neumann system on
the unit sphere $S^2=\{q\in{\mathbb R}^3 \mid q_1^2+q_2^2+q_3^2=1\}$
with the potential $U(q)=\frac12 ({\mathcal I} q,q)$,
\begin{equation}
\frac{d^2}{d\tau_1^2} q = -{\mathcal I} q + \lambda q,  \label{3.6}
\end{equation}
corresponding to the zero value of the integral
\begin{equation}
\left({\mathcal I}\left( \frac{d}{d\tau_1} q\times q\right), \frac{d}{d\tau_1}q\times q\right)
-\det{\mathcal I}\,  ({\mathcal I}^{-1} q,q).  \label{3.7}
\end{equation}
\end{thm}
\medskip

We emphasize that for $(\Omega,\gamma)\ne 0$, Theorem \ref{LR->N3} does not hold,
and in this case the procedure for integrating equations (\ref{3.1}), (\ref{la})
was indicated in \cite{Fed_nonholonomic}.
\medskip

\paragraph{Reduction.} The above relation between the LR system and the Neumann
system via the time reparameterization appears to be quite natural in view of the fact that
the Veselova system is  a Chaplygin system on the $SO(2)$--bundle
\begin{equation*}
\begin{array}{cccc}
SO(2) & \longrightarrow & SO(3) &  \\
&  & \downarrow & \pi \\
&  & S^2=SO(2)\backslash SO(3) &
\end{array},  \label{3.2}
\end{equation*}
where $SO(2)$ is the subgroup generated by rotation about the vector $\gamma$.
Indeed, the Lagrangian and the nonholonomic constraint (\ref{classical}) are
invariant  with respect to such rotations. Hence,  the Veselova system can be reduced to
the (co)tangent bundle of $S^2=\{q\in{\mathbb R}^3 \mid q_1^2+q_2^2+q_3^2=1\}$.

The momentum map $\Phi: TS^2 \to so(3)\cong \mathbb{\ R}^3$ is simply given by
$\Phi(q,\dot q)= \dot q\times q$, hence  the reduced Lagrangian is
$L(q,\dot q)=\frac12 \left( {\mathcal I} (\dot q\times q), \dot q\times q\right)-V(q)$.
Note that the reduced potential is given by the same function $V$, regarded
as a function of $q$ instead of $\gamma$.

Next, in view of the relation
$$
\pr_{g^{-1}\mathfrak{h} g} [\Phi(q,\dot q),\Phi(q, \xi)]
=(q, (\dot q\times q)\times (\xi\times q) )q
=\dot q\times \xi ,
$$
where $\xi=(\xi_1,\xi_2, \xi_3)^T$ is any tangent vector of $S^2$ at the point $q$,
the reduced Lagrange--d'Alembert equation (\ref{1.2}) takes the form
$$
\left( \frac{\partial L}{\partial q} - \frac{d}{dt}\frac{\partial L}{\partial \dot q}, \xi\right)
=\Psi(q,\dot q,\xi), \qquad \Psi= ({\mathcal I} (\dot q\times q), \dot q\times\xi).
$$
Now the reduced LR system on $T^* S^2$ can explicitly be written in terms of local
coordinates $q_1, q_2$
on $S^2$ and the corresponding momenta $p_1=\partial \tilde L/\partial {\dot q}_1$,
$p_2=\partial \tilde L/\partial {\dot q}_2$,
\begin{equation} \label{red2}
\frac{\partial \tilde L}{\partial q_k} - \frac{d}{dt} p_k
= \frac{\partial \tilde\Psi}{\partial \xi_k},
\qquad k=1,2,
\end{equation}
where $\tilde L, \tilde \Psi$ are obtained from $L(q,\dot q)$, $\Psi(q,\dot q,\xi)$
by the substitutions
$$
\dot q_3=-\frac{q_1\dot q_1+ q_2\dot q_2}{\sqrt{1-q_1^2-q_2^2}},
\quad
\xi_3 = - \frac{q_1\xi_1+ q_2\xi_2}{\sqrt{1-q_1^2-q_2^2}}\, .
$$

A direct (but tedious) calculation shows that  the reduced system (\ref{red2}) has
an invariant measure with density ${\mathcal N}(q)=1/\sqrt{(q,{\mathcal I}^{-1}q)}$.
(As was mentioned above, the latter does not depend on the choice of local
coordinates on $S^2$).

Since the reduced system is two-dimensional, Chaplygin's reducibility theorem
(item 2 of Theorem \ref{Th_reduce})
says that in the new time $d\tau=\mathcal{\ N} dt$ and new momenta
$\tilde p_k=\mathcal{\ N}p_k$,
$k=1,2$, equations  (\ref{red2})  transform to a Hamiltonian system.
Equivalently, the latter is described by
the following Lagrangian obtained from $L(q,\dot q)$,
\begin{equation} \label{3.3}
L^* (q,q^{\prime})=\frac1{2 (q,{\mathcal I}^{-1}q)}
\left({\cal I }(q'\times q), q'\times q) \right) -V(q), \qquad q'=\frac{dq}{d\tau}.
\end{equation}
For $V=0$, this is a Lagrangian of a geodesic flow on $S^2$.

\begin{thm}
The geodesic flow on $S^2$ with the metric
\begin{equation*}
(q, {\cal I}^{-1} q)^{-1} ds^2_{I,D}, \quad ds^2_{I,D}= {\det{\cal
I}} \left [  (dq, {\cal I}^{-1} dq)({\cal I}^{-1}q,q)-({\cal
I}^{-1} q,dq)^2\right]
\end{equation*}
obtained from (\ref{3.3}) is completely integrable. It has an
additional quadratic integral
\begin{equation*}
F^*_2(q,q^{\prime})= \frac1{2(q,{\mathcal I}^{-1}q)}
 \left( ( {\mathcal I} (q'\times q), q'\times q)- ( {\mathcal I} (q'\times q),q )^2\right),
\end{equation*}
which corresponds to the integral $F_2$ of the LR system (\ref{3.1}), (\ref{classical}).
\end{thm}

This theorem, as well as our observations on the reducibility of the Veselova system
to Hamiltonian form, is a part of a general integrability theorem for
a multi-dimensional Veselova system on the group $SO(n)$,
which we discuss in detail in Section 6.
In the rest of this section we only quote some specific properties of the 3-dimensional case.

\paragraph{The Veselova system on $SO(3)$ with integrable potentials.}
Classical integrable cases of the rigid body motion without constraints
were already used to produce integrable geodesic flows on the sphere
(see, e.g.,  \cite{BoKoFo}).
Namely, consider the Euler--Poisson equations of the motion of the rigid body with
tensor of inertia ${\cal J}$ and axisymmetric potential ${\cal V}(\gamma)$
\begin{equation}
{\mathcal J}\dot \Omega = {\mathcal J}\Omega \times \Omega
+ \gamma\times\frac{\partial {\cal V}}{\partial\gamma},  \quad
\dot \gamma = \gamma \times \Omega , \label{3.4}
\end{equation}
which always have first integrals
$$
i_1=(\gamma,\gamma)=1, \quad i_2=({\mathcal J}\Omega,\gamma), \quad
f_1=\frac12 ( {\mathcal J}\Omega,\Omega)+V(\gamma).
$$
In the Euler case $({\cal V}(\gamma)=0)$ there is an additional integral
$f_2=\frac12 ( {\mathcal J}\Omega, {\mathcal J}\Omega )$, and under the condition
$i_2=0$ and the substitution $\gamma =q$, equations (\ref{3.4}) define a geodesic flow on
the sphere $S^2$ with the metric
\begin{equation*}
ds^2_{J,P}= \frac{\det{\cal J}}{(q, {\mathcal J} q) }   (dq, {\cal J}^{-1} dq).
\end{equation*}

There is an interesting duality between integrable potentials and additional first integrals of
the Euler--Poisson equations and of the Veselova system.
 \begin{lem}
The Veselova system (\ref{classical}), (\ref{3.1}) with the potential
$V(\gamma)$ has an additional integral of the form $F=F_2 + F(\gamma)$
and therefore is integrable by the Euler--Jacobi theorem
if and only if the Euler--Poisson equations (\ref{3.4})
with inertia tensor ${\mathcal J}={\mathcal I}^{-1}$ and the potential
${\mathcal V}=F(\gamma)$ are integrable for $i_2=0$ due to the presence of the
extra integral $f_2+V(\gamma)$.
\end{lem}

\noindent{\it Proof.} Indeed, the necessary and sufficient
condition for equations (\ref{3.4}) with $i_2=0$ to have the
integral $\frac12 ({\mathcal J}\Omega,{\mathcal J}\Omega)+
V(\gamma)$ has the form
$$
\left[ \gamma\times \frac{\partial {\cal V}}{\partial\gamma}
+ {\cal J}^{-1}\left( \frac{\partial {V}}{\partial\gamma} \right)\times\gamma\right ]
\times\gamma =0\, .
$$
On the other hand, the system (\ref{classical}), (\ref{3.1}) has the integral
$ \frac12 ({\mathcal I}\Omega,{\mathcal I}\Omega)
- \frac12 ({\mathcal I}\Omega,\gamma)^2 + F(\gamma)$ if and only if
$$
\left[ \gamma\times \frac{\partial F}{\partial\gamma}
+ {\cal I}\left( \frac{\partial {V}}{\partial\gamma} \right)\times\gamma\right ]
\times\gamma =0\, .
$$
Since we set ${\mathcal J}^{-1}={\mathcal I}$, $F(\gamma)={\cal V}(\gamma)$, both
conditions are equivalent, which proves the lemma.
\medskip

Some integrable  polynomial potentials for the Euler--Poisson equations are
given in \cite{Bo}. In a similar way, one can construct  integrable polynomial potentials
(or Laurent polynomial potentials, such as given in \cite{Dr}) for the
Veselova system. For example, the following proposition holds.

\begin{prop} Let ${\mathcal I}=\mbox{diag }(I_1, I_2, I_3)$.
The Veselova system (\ref{classical}), (\ref{3.1}) with potential
\begin{equation}
V(\gamma)= \alpha_1\left( ({\mathcal I}^2\gamma,\gamma) -
({\mathcal I}\gamma,\gamma)^2\right)+ \alpha_2 ({\cal I}\gamma,\gamma)+
\frac{\alpha_3}{\gamma_1^2}+\frac{\alpha_4}{\gamma_2^2}
+ \frac{\alpha_5}{\gamma_3^2},  \label{3.5}
\end{equation}
$\alpha_1,\dots, \alpha_5$ being arbitrary constants, is solvable by quadratures.
The additional integral is:
\begin{eqnarray*}
F&=&\frac12 ({\cal I}\Omega,\Omega) - \frac12
({\cal I}\Omega,\gamma)^2+ \alpha_1\det{\cal I} ( {\cal I}\gamma,\gamma)
({\cal I}^{-1}\gamma,\gamma) -\alpha_2\det{\cal I}
( {\cal I}^{-1}\gamma,\gamma) \\
&&+\alpha_3\left(I_2\frac{\gamma_2^2}{\gamma_1^2}+ I_3\frac{\gamma_3^2}
{\gamma_1^2}\right) +\alpha_4\left(I_3\frac{\gamma_3^2}{\gamma_2^2}
+ I_1\frac{\gamma_1^2}{\gamma_2^2}\right)
+\alpha_5\left(I_1\frac{\gamma_1^2}{\gamma_3^2}
+ I_2\frac{\gamma_2^2}{\gamma_3^2}\right).
\end{eqnarray*}
\end{prop}

Note that the integrability of the Veselova system with the Clebsch potential
$\alpha ({\cal I}\gamma,\gamma)$ was already shown in \cite{VeVe1, VeVe2}.

\section{Nonholonomic LR systems on $SO(n)$ and their reductions to Stiefel varieties}

Now we proceed to a generalization of the Veselova system,
which describes the motion of an $n$-dimensional rigid body with a fixed point,
that is, the motion on the Lie group $SO(n)$,
with certain right-invariant nonholonomic constraints.

For a path $g(t)\in SO(n)$,  the angular velocity of the body is given by
the left-trivialization $\omega(t)=g^{-1}\cdot g(t)\in so(n)$.  The
matrix $g\in SO(n)$ maps a coordinate system fixed in the body to
a coordinate system fixed in the space.
Therefore, if $e_1=(e_{11},\dots,e_{1n})^T, \dots,e_n=(e_{n1},\dots,e_{nn})^T$
is an orthogonal frame of unit vectors fixed in the space and
{\it regarded in the moving frame}, we have
\begin{gather*}
E_1=g\cdot e_1,\; \dots,\;  E_n=g\cdot e_n,
\end{gather*}
where $E_1=(1,0,\dots,0)^T,\; \dots,\; E_n=(0,\dots,0,1)^T$. From
the conditions $0=\dot E_i=\dot g \cdot e_i+g\cdot \dot e_i$, we
find that the vectors $e_1,\dots,e_n$ satisfy the Poisson
equations
\begin{equation} \label{Po}
\dot e_i= - \omega e_i, \qquad  i=1,\dots,n.
\end{equation}

Below we use the convention $x\wedge y=x\otimes
y-y\otimes x=x\cdot y^T-y\cdot x^T$. Also now $\langle\cdot,\cdot\rangle$ denotes the Killing
metric on $so(n)$, $\langle X,Y\rangle=-\frac12 tr(XY)$, $X,Y\in so(n)$.
The left-invariant metric on $SO(n)$ is given by a non-degenerate inertia operator
${\mathcal I}\,:\, so(n)\to so(n)$ and the Lagrangian of the free motion of the body is
$l=\frac12\langle {\mathcal I}\omega,\omega\rangle$.
\medskip

\paragraph{Right-invariant constraints on $SO(n)$.}
What form may have a multi-dimensional analog of the classical constraint (\ref{classical})?
To answer this question, we first note that instead of rotations {\it about an axis}
in the classical mechanics, in the $n$-dimensional case there are
infinitesimal rotations in two-dimensional planes spanned by the
basis vectors $e_{i},e_{j}$,  $i,j=1,\ldots,n.$
Suppose, without loss of generality, that in the three-dimensional case $\gamma=e_1$.
Then condition (\ref{classical}) can be redefined as follows: only infinitesimal rotations
in the planes span$(e_{1}, e_{2})$ and span$(e_{1}, e_{3})$ are allowed. Hence,
it is natural to define its $n$-dimensional analog
as follows: only infinitesimal rotations in the fixed 2-planes
spanned by $(e_{1}, e_{2}), \dots, (e_{1}, e_{n})$ (i.e., in the planes containing
the vector $e_{1})$ are allowed. This implies the constraints
\begin{equation}
\langle \omega, e_i\wedge e_j\rangle=0, \quad 2\le i<j\le n.  \label{44}
\end{equation}

Following \cite{FeKo2}, one can relax these constraints by assuming that
the angular velocity matrix {\it in the space} has the following structure
\begin{equation*}
\tilde\omega= g \omega g^{-1} =
\begin{pmatrix}
0 & \cdots & \omega_{1r} & \cdots & \omega_{1n} \\
\vdots & \ddots & \vdots &  & \vdots \\
-\omega_{1r} & \cdots & 0 & \cdots & \omega_{rn} \\
\vdots &  & \vdots & \mathbf{O} &  \\
- \omega_{1n} & \cdots & - \omega_{rn} &  &
\end{pmatrix},
\end{equation*}
where $\mathbf{O}$ is the zero $(n-r)\times (n-r)$ matrix.

Equivalently,
consider the right--invariant distribution $D_r$ on $TSO(n)$
whose restriction to the algebra $so(n)$ is given by
\begin{equation*}
\mathfrak{d}=\Span\{E_j \wedge E_k, \; k=1,\dots, r, \; j=1,\dots, n\},
\end{equation*}
where $E_i \wedge E_j$ form the basis in $so(n)$.
Since $e_i \wedge e_j=g^{-1} \cdot E_i \wedge E_j \cdot g$, we find that
the constraints are defined by relations
\begin{gather}
\omega \in {\cal D}_r= g^{-1}\cdot \mathfrak{d} \cdot g
=\Span\{e_1 \wedge e_i, \dots, e_r \wedge e_i, \;
1\le i \le n\} , \nonumber \\
\mbox{that is } \quad \langle \omega, e_p\wedge e_q\rangle=0, \quad r< p<q\le n.
\label{4.1}
\end{gather}

The LR system on the right-invariant distribution $D_r \subset T\,SO(n)$
can be described by the Euler--Poincar\'e  equations (\ref{EPS})
on the space $so(n)\times SO(n)$ with indefinite multipliers $\lambda_{pq}$,
\begin{align}
\dot {\mathcal I}\omega +[\omega,{\mathcal I}\omega ]
& =\sum_{r<p<q\le n}\lambda_{pq} \, e_p\wedge e_q , \nonumber \\
\dot{e}_{i}+\omega e_{i} &=0 , \qquad i=1,\dots, n . \label{nh3.46}
\end{align}
Here the components of the vectors $e_1,\dots, e_n$ play the role of redundant
coordinates on $SO(n)$.
For $n=3, r=1$, after identification of the Lie algebras
$(\mathbb{R}^3,\times)$ and $(so(3),[\cdot,\cdot])$ and setting
$\omega_{ij}=\varepsilon_{ijk}\Omega_k$, $e_1=\gamma$,
this becomes the Veselova system (\ref{3.1}) with $V=0$.

By analogy with (\ref{3.1}), we will call $(SO(n), l, D_r)$
\textit{a multidimensional Veselova system}.

Differentiating  (\ref{4.1}), from (\ref{nh3.46}) one can obtain a system of linear
equations for the determination of the multipliers in terms of the components of
$\dot \omega, \omega$, and $e_i$. Thus, (\ref{nh3.46})  contains a closed system of differential
equations on the space $(\omega_{ij}, e_{r+1},\dots, e_n)$. The latter system has first integrals
\begin{equation}
\langle\omega, e_p\wedge e_q\rangle= w_{pq}, \quad   w_{pq}=\mbox{const},
\qquad r< p<q\le n \label{4.1a}
\end{equation}
and our LR system on $D_r \subset T\,SO(n)$ is the restriction of (\ref{nh3.46}) onto
the level variety $w_{pq}=0$.

As follows from Theorem \ref{measure1}, the system (\ref{nh3.46})
has an invariant measure with density
\begin{gather*}
\mu =\sqrt{\det \left( {\cal I}^{-1}|_{\bot{\cal D}_r }\right) } =
\sqrt{|\langle  e_p\wedge e_q, {\mathcal I}^{-1}(e_s \wedge e_l) \rangle | } , \\
r<p<q\le n, \quad r<s<l\le n,
\end{gather*}
where ${\bot\cal D}_r \subset so(n)$ is the orthogonal complement of ${\cal D}_r$
with respect to the metric $\langle\cdot,\cdot\rangle$ and
${\mathcal I}^{-1}|_{\bot{\cal D}_r}$ is the restriction of the inertia tensor to
$\bot{\cal D}_r $.

In the case of the Veselova system on $TSO(3)$ after identifying
$e_1=e_2\times e_3$ with $\gamma\in {\mathbb R}^3$
the above expression reduces to the known form $\sqrt{(\gamma, I^{-1}\gamma)}$.

In practice, for a large dimension $n$ and small $r$, the number of constraints
(\ref{4.1})  is large,
which leads to rather tedious expressions for the explicit form of the system and the density of
its invariant measure.
In this case one can make use of the alternative momentum description
(system (\ref{mom_eq})). Namely, in view of the matrix representation
\begin{gather*}
\forall \; X\in so^*(n), \quad \pr_{{\cal D}_r } (X)= \Gamma X+ X\Gamma -\Gamma X\Gamma , \\
\Gamma  = e_1\otimes e_1+\cdots +  e_r\otimes e_r \ ,
\end{gather*}
system (\ref{mom_eq1}) takes the following form:
\begin{align}
\dot {\cal M} &= [{\cal M},\omega], \label{4.1c} \\
{\cal M} &=\pr_{{\cal D}_r} ({\mathcal I}\omega)+ \pr_{\bot{\cal D}_r}\omega
\nonumber \\
& \equiv \omega + ({\mathcal I}\omega -\omega )\Gamma +\Gamma  ({\mathcal I}\omega -\omega )
- \Gamma  ({\mathcal I}\omega -\omega ) \Gamma \,  . \label{4.2}
\end{align}
The map $\omega \to {\cal M}$ given by (\ref{4.2}) is nondegenerate.
As a result, equations (\ref{4.1c}), (\ref{4.2}) together with the Poisson equations
(\ref{Po}), which are equivalent to
\begin{equation}
\dot \Gamma= [\Gamma, \omega],
\label{4.2a}
\end{equation}
represent a closed system of differential equations on the space $(\omega, e_1,\dots, e_r)$
or on the space $({\cal M}, e_1,\dots, e_r)$.

In the classical case of $n=3, r=1$, passing to the vector variables
$\gamma, \Omega$, ${\mathfrak m}$, where ${\cal M}_{ij}=\varepsilon_{ijk}{\mathfrak m}_k$,
we obtain
$$
{\mathfrak m} = {\mathcal I} \Omega -({\mathcal I}\Omega,\gamma)\gamma + (\Omega, \gamma)\gamma,
\quad \mbox{and } \quad
{\mathcal I} \Omega = {\mathfrak m} - \frac{({\mathfrak m}, {\cal I}^{-1}\gamma)-
({\mathfrak m},\gamma)} {(\gamma, {\cal I}^{-1}\gamma)}.
$$
Substituing this into (\ref{4.1c}) yields explicit equations in terms of $\Omega, \gamma$,
which again describe the Veselova LR system (\ref{3.1}), (\ref{la}) with $V=0$.

As follows from the structure of the system (\ref{4.1c}), (\ref{4.2a}),
it possesses a family of integrals given by nonzero coefficients of the
following polynomial in $\lambda$
$$
\mbox{tr }\, ({\cal M}+ \lambda\Gamma)^k , \qquad k \in{\mathbb N}.
$$
In addition, it has the invariant variety defined by the condition
\begin{gather}
\label{D}
{\mathcal M} \, \wedge e_1 \wedge \cdots
\wedge  e_r \equiv \omega  \, \wedge e_1 \wedge \cdots \wedge e_r  = 0,
\end{gather}
where ${\mathcal M}, {\omega}$ are considered as 2-forms and
$e_k$ as 1-forms in the same Euclidean space ${\mathbb R}^n$.
This gives a set of scalar conditions on the components
of ${\mathcal M}$ or $\omega$,  which describes
the linear subspace ${\cal D}_r = g^{-1}\cdot \mathfrak{d} \cdot g\subset so(n)$.
Hence, among conditions (\ref{D}) only $(n-r)(n-r-1)/2$ are independent.

Next, according to Theorem \ref{IM2},
the LR system (\ref{4.1c}), (\ref{4.2}), (\ref{4.2a})
possesses an invariant measure with dual density
\begin{gather}
\Theta=\tilde\mu \, d \omega \wedge d e_1 \wedge \cdots \wedge d e_r =
{\tilde \mu}^{-1} \, d {\mathcal M} \wedge d e_1 \wedge \cdots \wedge d e_r
\nonumber \\
\tilde \mu= \sqrt{ \det ({\mathcal I}|_{{\cal D}_r}) }=
\sqrt{|\langle  e_i\wedge e_p, {\mathcal I} (e_j \wedge e_q) \rangle | } ,
\label{IM3} \\
1 \le p<q\le r, \quad 1\le i<j\le n, \nonumber
\end{gather}
where ${\mathcal I}|_{{\cal D}_r}$ is the restriction of the inertia tensor to
 the subspace ${\cal D}_r \subset so(n)$.

\begin{remark} {\rm Since ${\cal D}_r$ is invariant under the action of
$SO(n-r)$ on the linear space spanned by the vectors
$e_{r+1},\dots, e_n$,  expression (\ref{IM3}), in fact, does not
depend explicitly on the components of these vectors. Moreover,
${\cal D}_r$ is also invariant under the $SO(r)$-action on the
space span $(e_1,\dots, e_r)$. As a result, the above density
depends explicitly only on coordinates on the Grassmannian
$G(r,n)=SO(n)/(SO(n-r)\times SO(r))$, that is, on the {\it
Pl\"ucker coordinates\/} of the $r$-form $e_1 \wedge \cdots \wedge
e_r$, which are invariants of the above actions. A simplified
expression for the density depends on the choice of $\cal
I$.}\end{remark}

\paragraph{The special inertia tensor.}
It appears that for some special inertia tensors, the density (\ref{IM3})
takes an especially simple form, which we shall use in the sequel.
Suppose that the operator ${\mathcal I}$ is defined by a diagonal matrix
$A=diag(A_1,\dots,A_n)$ in the following way
\begin{equation}
{\mathcal I} (E_i\wedge E_j)=\frac{A_i A_j}{\det A} E_i\wedge E_j  . \label{4.5}
\end{equation}
Notice that for $n=3$ this corresponds to the well known three-dimensional vector formula
${\cal I}(x\times y)=\frac{1}{\det A} Ax \times Ay$, $A={\cal I}^{-1}$ and thus in this case
defines a {\it generic}  inertia tensor.

\begin{thm} \label{IM_A}
 Under the above choice of ${\mathcal I}$,
\begin{equation} \label{det_sp}
 \det ({\mathcal I}|_{{\cal D}_r})={\mathcal P}_{n,r}
=(\det A)^\rho \left[ \sum_{I} A_{i_1} \cdots A_{i_r} (e_1 \wedge
\cdots \wedge e_r)^2_{I} \right]^{n-r-1},
\end{equation}
where $\rho$ is an integer constant, the summation is over all $r$-tuples \\
$I=\{1\le i_1 < \cdots < i_r \le n\}$, and $(e_1 \wedge \cdots \wedge e_r)_{I}$
are the {\it Pl\"ucker coordinates\/} of the $r$-form $e_1 \wedge \cdots \wedge e_r$.
\end{thm}

\noindent{\it Proof.} It is more convenient to calculate first the dual determinant
$|{\mathcal I^{-1}}|_{\bot{\cal D}_r}|$, which can be represented in the form
\begin{gather}
 \bigg |\det A (e_{p}, A^{-1} e_{s}) (e_{q}, A^{-1} e_{l}) - \det A  (e_{p}, A^{-1} e_{l}) (e_{q}, A^{-1} e_{s})
\bigg | \, ,   \label{exp} \\
r<p<q\le n, \quad r<s<l\le n . \nonumber
\end{gather}
Since we deal with purely algebraic expressions, in this proof we can regard
$e_{r+1},\dots, e_n$ as vectors in the complex space ${\mathbb C}^n$.
Next, since the action of $SO(n-r)$ on the linear
space ${\bar\Lambda} \subset {\mathbb C}^n$ spanned by $e_{r+1},\dots, e_n$ does
not change $\bot{\cal D}_r \subset \wedge^2{\mathbb C}^n$,
the above determinant  must depend only on the Pl\"ucker coordinates
$$
(e_{r+1} \wedge \cdots \wedge e_n)_J,  \quad J=\{j_1, \dots, j_{n-r}\}, \quad
1\le j_1 < \cdots < j_{n-r} \le n .
$$
In view of dimension and the structure of the determinant (\ref{exp}),
it is a homogeneous polynomial in the components of $e_p$ of degree
$$
4\cdot\mbox{ dim } SO(n-r)=2(n-r)(n-r-1).
$$
Hence, it is a homogeneous polynomial of degree $2(n-r-1)$ in
$(e_{r+1} \wedge \cdots \wedge e_n)_J$.

Suppose that the linear space ${\bar\Lambda}$ is tangent to a (possibly imaginary) cone
${\mathcal K}=\{(X, A^{-1} X)=0 \} \subset {\mathbb C}^n$ and, without loss of generality,
assume that $e_n$ is directed along the tangent line ${\bar\Lambda}\cap {\mathcal K}$. Then
$(e_n, A^{-1} e_p)=0$ for $p=r+1,\dots,n$, and in this case the last $n-r-1$ rows and
columns of the determinant (\ref{exp}), and therefore the determinant itself, vanish.

On the other hand, the condition for ${\bar\Lambda}$ to be tangent to ${\mathcal K}$ has the form
$$
\det (A^{-1}|_{\bar\Lambda}) = \left |
\begin{matrix}  (e_{r+1},A^{-1} e_{r+1}) & \cdots &  (e_{r+1},A^{-1} e_n) \\
                                                           \vdots  &   & \vdots \\
             (e_n, A^{-1} e_{r+1}) & \cdots &  (e_n, A^{-1} e_n) \end{matrix} \right|=0 \, ,
$$
where $A^{-1}|_{\bar\Lambda}$ is the restriction of the quadratic form $A$ onto $\bar\Lambda$.
Expanding the latter determinant we see that it equals
$\sum_{J} A^{-1}_{i_1} \cdots A^{-1}_{i_r} (e_{r+1} \wedge\cdots \wedge e_n)^2_J$,
thus it is a quadratic polynomial in the above Pl\"ucker coordinates.

Combining our considerations, we see that when ${\bar\Lambda}$ is tangent to
${\mathcal K}$, the matrix $A^{-1}|_{\bar\Lambda}$ has corank 1, whereas
the matrix ${\mathcal I^{-1}}|_{\bot{\cal D}_r}$ has corank  $(n-r-1)$.
Hence, the determinant (\ref{exp}) is divisible by the $(n-r-1)$-th power of
$\det (A^{-1}|_{\bar\Lambda})$, a homogeneous polynomial
of degree $2(n-r-1)$ in the coordinates $(e_{r+1} \wedge\cdots \wedge e_n)_J$.
Thus the quotient of the determinant and the polynomial has zero degree in these coordinates.
Since the quotient  cannot have poles, it is a constant.
An additional study of (\ref{exp}) shows that this constant
must be a positive power of $\det A$. As a result,
$$
\det ({\mathcal I^{-1}}|_{\bot{\cal D}_r })
=(\det A)^{\rho_1} \left[ \sum_{J} A^{-1}_{j_1} \cdots A^{-1}_{j_{n-r}}
(e_{r+1} \wedge\cdots \wedge e_n)^2_J\right ]^{n-r-1}, \quad \rho_1\in{\mathbb N}.
$$

Now, in order to obtain the desired expression (\ref{det_sp}),
we use the relations (\ref{duality}) with $g^{-1}{\mathfrak d}g ={\cal D}_r$,
$g^{-1}{\mathfrak h}g =\bot {\cal D}_r$, and
$(e_{r+1} \wedge\cdots \wedge e_n)_J^2= (e_1 \wedge\cdots \wedge e_r)^2_{I}$,
where $I$ and $J$ are complimentary multi-indices in the sense that
$\{ i_1, \dots, i_r, \, j_1, \dots, j_{n-r}\}$ is a permutation of $\{1,\dots, n\}$.
This, together with the fact that $\det {\mathcal I}$ is a power of $\det A$, proves the theorem.
\medskip

From Theorems \ref{IM1}, \ref{IM_A} we get

\begin{cor} Under the condition (\ref{4.5}),
the LR system \textup{(\ref{4.1c})--(\ref{4.2a})}
has an invariant measure
\begin{equation} \label{true}
\left[ \sum_{I} A_{i_1} \cdots A_{i_r} (e_1 \wedge
\cdots \wedge e_r)^2_{I} \right]^{-(n-r-1)/2}\, d{\cal M}\wedge d\,e_1\wedge\cdots \wedge d\,e_r\, .
\end{equation}
In the particular case of $r=1$, the density $\tilde\mu$ in (\ref{IM3}) is proportional to
$(e_1, A e_1)^{(n-2)/2}$.
\end{cor}

As follows from  (\ref{det_sp}) or (\ref{true}), in the opposite extreme case $r=n-1$
(no constraints) ${\mathcal P}_{n,r}$  and $\tilde\mu$ are just constants, as expected.
\medskip

\paragraph{Reduction to Stiefel varieties.}
Now we notice that in the case of the constraints (\ref{4.1}),
the orthogonal complement $\mathfrak{h}$ of $\mathfrak{d}$ is a Lie algebra, namely
$$
\mathfrak{h}=\Span\{E_p \wedge E_q, \; r < p<q \le n\}\cong so(n-r).
$$
Therefore, according to the observations of Section 2,
the multidimensional Veselova system can be treated as a
generalized Chaplygin system on the principal bundle
\begin{equation}
\begin{array}{cccc}
SO(n-r) & \longrightarrow & SO(n) &  \\
&  & \downarrow & \pi \\
&  &  V(r,n)=SO(n-r)\backslash SO(n) &
\end{array},
\end{equation}
where $V(r,n)$ is the Stiefel variety. It can be regarded as
the variety of ordered sets of $r$ orthogonal unit  vectors
$e_1,\dots, e_r$ in ${\mathbb R}^n$ (${\mathbb C}^n$),  or,
equivalently, the set of $r \times n$ matrices $ {\mathcal X}
=(e_1 \cdots e_r)$ satisfying ${\mathcal X}^T {\mathcal X}={\bf
I}_r$, where ${\bf I}_r$ is the $r\times r$ unit matrix. Thus
$V(r,n)$ is a smooth variety of dimension $N=rn-r(r+1)/2$ (see e.g.,
\cite{Sovr_Geom}), and the components of $ {\mathcal X}$ are
redundant coordinates on it.

The nonholonomic right-invariant distribution $D_r$ is orthogonal
to the leaf of the action of $SO(n-r)$ with respect to the bi-invariant metric on $SO(n)$.

The tangent bundle $TV(r,n)$ is the set of pairs
${\mathcal X}, \dot {\mathcal X}$ subject to the constraints
\begin{equation} \label{cond_X}
{\mathcal X}^T {\mathcal X}={\bf I}_r, \quad {\mathcal X}^T
\dot{\mathcal X} + \dot{\mathcal X}^T {\mathcal X} =0,
\end{equation}
which give $r(r+1)$ independent scalar constraints.

\begin{lem}
The momentum map  $\Phi\, :\, TV(r,n) \to so(n)$  is given by
\begin{align}
\omega & = \Phi( {\mathcal X} ,\dot {\mathcal X} )
=  {\mathcal X}  \dot {\mathcal X}^T-  \dot {\mathcal X}  {\mathcal X}^T
+\frac 12 {\mathcal X} [ {\mathcal X}^T \dot  {\mathcal X}
-  \dot {\mathcal X}^T {\mathcal X}] {\mathcal X}^T  \nonumber \\
& \equiv e_1 \wedge \dot e_1+\cdots +  e_r \wedge \dot e_r +\frac 12
\sum_{1\le\alpha<\beta\le r} \left[  (e_\alpha, \dot e_\beta)-
(\dot e_\alpha, e_\beta)\right] \, e_\alpha \wedge e_\beta .
\label{4.3}
\end{align}
\end{lem}

Indeed, $\Phi(\mathcal X,\dot{\mathcal X})^T =-\Phi(\mathcal X,\dot{\mathcal X})$
and therefore $\Phi(\mathcal X,\dot{\mathcal X})\in so(n)$.
Taking into account constraints (\ref{cond_X}), we obtain
$$
-\Phi( {\mathcal X} ,\dot {\mathcal X}){\mathcal X}
= \dot {\mathcal X}- {\mathcal X} \dot {\mathcal X}^T {\mathcal X}
- \frac 12 {\cal X}( {\mathcal X}^T \dot {\mathcal X}
 - \dot{\mathcal X}^T {\mathcal X}) \, ,
$$
which implies the Poisson equations for $e_i$,
\begin{equation} \label{Poisson}
\dot{\mathcal X}=-\omega {\mathcal X}.
\end{equation}
On the other hand, putting $\dot{\mathcal X}=-\omega {\mathcal X}$
into $\Phi({\mathcal X} ,\dot {\mathcal X})$, we get
$$
\Phi=\omega\Gamma + \Gamma \omega-\Gamma\omega\Gamma = \pr_{{\cal D}_r} (\omega), \qquad
\Gamma  = e_1\otimes e_1+\cdots +  e_r\otimes e_r .
$$
Hence $\Phi ({\mathcal X} ,\dot {\mathcal X}) \in {\cal D}_r$ and formula (\ref{4.3})
describes the momentum mapping.
\medskip

The reduced Lagrangian $L({\mathcal X}, \dot{\mathcal X})$ takes the form
$$
L =\frac 12   \langle {\cal I} \Phi({\mathcal X},\dot {\mathcal X}),
\Phi({\mathcal X},\dot {\mathcal X})\rangle =
- \frac 1{4} \mbox{\bf tr}\left({\cal I} \Phi({\mathcal X},\dot {\mathcal X})\circ
\Phi({\mathcal X},\dot {\mathcal X})\right).
$$
Then we introduce the $r \times n$ momentum matrix
\begin{equation}
{\mathcal P}_{is}=\partial L({\mathcal X}, \dot{\mathcal X}) /\partial
\dot{\mathcal X}_{is}\, .
\label{dotX->P}
\end{equation}
Since the Lagrangian is degenerate in the redundant velocities
$\dot{\mathcal X}_{is}$, from this relation
one cannot express $\dot{\mathcal X}$ in terms of ${\mathcal X}, {\mathcal P}$  uniquely.
On the other hand, the cotangent bundle $T^*V(r,n)$ can be realized as the set of pairs
${\mathcal X}, {\mathcal P}$ satisfying the constraints
\begin{equation} \label{condX}
{\mathcal X}^T {\mathcal X}
={\bf I}_r, \quad {\mathcal X}^T {\mathcal P} + {\mathcal P}^T {\mathcal X} =0\, .
\end{equation}
The corresponding symplectic structure $\varOmega$ on $T^*V(r,n)$ is just the restriction of the
canonical 2-form on the ambient space ${\mathbb R}^{2nr}=({\cal X}, {\cal P})$,
$$
\sum_{i=1}^n \sum_{s=1}^r d{\cal P}_{is}\wedge \, d{\cal X}_{is}\, .
$$
Under conditions (\ref{condX}), relation (\ref{dotX->P}) can be uniquely inverted,
i.e., one gets $\dot{\mathcal X}=\dot{\mathcal X}(\mathcal X,\mathcal P)$
(for $r=1$ see the section below).

Next, according to the definition of the reduced momentum map in (\ref{*}) and in
view of (\ref{4.3}), (\ref{dotX->P}),
the map $\Phi^*\, : T^*_{\cal X} V(r,n)\to so^*(n)$ has the form
\begin{equation} \label{mom^*}
 \Phi^*({\mathcal X}, {\mathcal P})={\mathcal I}\omega |_{{\cal D}_r}
=  {\mathcal X} {\mathcal P}^T-  {\mathcal P} {\mathcal X}^T
+\frac 12 {\mathcal X} [ {\mathcal X}^T {\mathcal P}
-  {\mathcal P}^T {\mathcal X}] {\mathcal X}^T .
\end{equation}
It establishes a bijection between the linear subspace
${\cal D}_r \subset so^*(n)=\{{\cal M}\}$ and the cotangent space $T^*_{\cal X} V(r,n)$.

\begin{thm} \label{rest_St}
The reduced LR system on $T^*V(r,n)$ is the restriction of  the following system on
the space $({\mathcal X}, {\mathcal P})$,
\begin{equation} \label{dot_P}
\dot{\mathcal X}=-\omega ({\cal X}, {\cal P})\, {\mathcal X}, \quad
\dot{\mathcal P}= -\omega ({\cal X}, {\cal P})\,   {\mathcal P},
\end{equation}
where $\omega ({\cal X}, {\cal P})=\Phi({\cal X}, \dot{\cal X}({\cal X}, {\cal P}))$.
\end{thm}

\noindent{\it Proof.}
Substituting the expression (\ref{mom^*}) into
 the reduced momentum equation (\ref{**}), differentiating its left hand side,
then taking into account the Poisson equations (\ref{Poisson})
and the conditions (\ref{condX}), we obtain
$$
{\cal X}\dot{\cal P}^T -\dot{\cal P}{\cal X}^T
-{\cal X}(\dot{\cal P}^T{\cal X}-{\cal P}^T\omega{\cal X}){\cal X}^T
= {\cal X}{\cal P}^T\omega +\omega {\cal P}{\cal X}^T .
$$
Multiplying  both sides of this relation from the left by ${\mathcal X}^T$ and
from the right by ${\mathcal X}$, then using again the conditions (\ref{condX}),
we arrive at the equation
$$
- {\cal X}^T\dot{\cal P}={\cal X}^T\omega{\cal P},
$$
which implies that $\dot{\cal P}=-\omega P$, i.e., the second equation in
(\ref{dot_P}).
The first equation in this system is just a repetition of (\ref{Poisson}).
The theorem is proved.
\medskip

According to (\ref{dot_P}), apart from the energy integral,
the reduced flow on $T^* V(r,n)$ possesses matrix momentum integral
${\cal P}^T{\cal P}$.

Notice that the form of equations (\ref{dot_P}) is similar to those describing
geodesic flows on Stiefel and Grassmann varieties (see \cite{Bl_Br_cr, BJ}) and on
other homogeneous spaces (\cite{Th, Br, BJ2}).
However, our system is not Hamiltonian with respect to the symplectic structure
$\varOmega$.

\paragraph{Reduced invariant measure.}
The phase space $({\cal M}, e_1,\dots, e_r)$ of the LR system
(\ref{4.1c}), (\ref{4.2}), (\ref{4.2a}) has the structure of the dual to
the semi-direct Lie algebra product
$$
so(n) \ltimes ( \underbrace{ {\mathbb R}^n\times \cdots
\times  {\mathbb R}^n}_{r \mbox{ times}} )
$$
and carries the corresponding Lie--Poisson structure $\{\cdot , \cdot \}_s$.
(For $r=1$ this is just the Lie--Poisson bracket on the coalgebra $e^*(n)$.)
This Poisson structure is degenerate, and the subvariety
${\cal O}_r\subset ({\cal M}, e_1,\dots, e_r)$ defined
by the constraints (\ref{D}) and the conditions
$$
\phi_{kl}=(e_k, e_l)=\delta_{kl}, \qquad k,l=1,\dots,r
$$
is its symplectic leaf of dimension $2N=2\textup{ dim } V(r,n)=rn-r(r+1)/2$:
the restriction of $\{\cdot, \cdot \}_s$ onto ${\cal O}_r$ is nondegenerate.
The variety ${\cal O}_r$ is a (generally singular) orbit of coadjoint action
of the semi-direct group product
$SO(n) \ltimes ({\mathbb R}^n\times \cdots \times  {\mathbb R}^n)$.

Let $\varSigma$ be the corresponding symplectic structure on ${\cal O}_r$.
By construction, the extended momentum map
$$
\tilde{\Phi}^*\,:\, T^* V(r,n) \to {\cal O}_r , \quad
({\mathcal X}, {\mathcal P}) \to (e_1,\dots, e_r, \Phi^*({\mathcal X}, {\mathcal P}))
$$
preserves the Poisson structure, hence it is a symplectomorphism:
the 2-form  $\varSigma$ passes to the symplectic structure
$\varOmega$ on  $T^* V(r,n)$.
Thus, $\varSigma^N$, as a volume form on ${\cal O}_r$, transforms
to the canonical volume form on  $T^* V(r,n)$.
\medskip

As we know from Theorem 3.3, a reduced LR system always has an invariant measure.
Using the above property of $\tilde{\Phi}^*$, the measure in our example can be
written explicitly.

\begin{thm} \label{mes} The reduced LR flow on $T^* V(r,n)$ has invariant measure
$$
1/\sqrt{ \det ({\mathcal I}|_{{\cal D}_r})  } \, \varOmega^N, \qquad
N=\textup{dim }  V(r,n)=rn-r(r+1)/2.
$$
\end{thm}

Notice that the density of this measure coincides with that of the invariant measure $\Theta$
of the LR system  (\ref{4.1c})--(\ref{4.2a})  in the coordinates
$({\cal M}, e_1,\dots, e_r)$. In particular, for the special inertia tensor (\ref{4.5})
the density is the same as in (\ref{true}).
\medskip

 \noindent{\it Sketch of proof of Theorem} \ref{mes}. Let
$$
\psi_1 ({\cal M},e), \dots, \psi_{m}({\cal M},e), \qquad  m=(n-r)(n-r-1)/2
$$
be any independent linear combinations of the constraint functions defined by (\ref{D}).
Then, at points of ${\cal O}_r\subset ({\cal M}, e_1,\dots, e_r)$,
\begin{align}
d{\cal M}_{12} \wedge & \cdots\wedge d{\cal M}_{n-1,n}
\wedge d\,e_{11}\wedge\cdots \wedge d\,e_{rn} \nonumber \\
& =  \xi(e_1, \dots, e_r)
 \cdot d\psi_1\wedge\cdots\wedge d\psi_m \wedge \varSigma^N
\prod_{1\le k\le l\le r} \wedge \, d\phi_{kl},  \label{forms}
\end{align}
where, as above, $\phi_{kl}=(e_k, e_l)$ and
$\xi(e_1, \dots, e_r)$ is a certain nonvanishing function.
Let $X_{{\cal M}_{ij}}, X_{e_{is}}$ be the basis vectors in the phase space of the
LR system. The function $\xi$ can be found by inserting  the polyvector
$$
X_{{\cal M}_{12}} \wedge \cdots\wedge X_{{\cal M}_{n-1,n}}\wedge
X_{e_{11}}\wedge \cdots \wedge X_{e_{nr}}
$$
into the left and right hand sides of (\ref{forms}) and taking
into account the relations
\begin{gather*}
\varSigma (X_{{\cal M}_{ij} }, X_{{\cal M}_{pq} }) = \{ {\cal
M}_{ij}, {\cal M}_{pq} \}, \quad \varSigma ( X_{ {\cal M}_{ij} },
X_{e_{p k}})
= \{ {\cal M}_{ij}, e_{p k} \},  \\
\varSigma (X_{e_{p k}}  , X_{e_{q l}} )= \{e_{p k}, e_{q l}\}= 0.
\end{gather*}

One can always choose such a basis of functions $\psi_k ({\cal M},e)$ that $\xi$
becomes a constant on the whole orbit  ${\cal O}_r$.
Moreover, for this basis, the time derivative with respect to the flow
(\ref{4.1c})--(\ref{4.2a}) has the form
$\dot \psi_k =  \sum_{s=1}^m \varkappa_{ks} \psi_s$ where the functions $\varkappa_{ks}$
satisfy the conditions $\varkappa_{kk}=0$ on ${\cal O}_r$.

Now let ${\cal L}_v=d\, i_v+ i_v\,  d$ denote  the Lie derivative
in the space $({\cal M}, e_1,\dots, e_r)$ with respect to this
flow, $i_{v}$ being  the interior product corresponding to the
flow. Since the functions $\phi_{kl}$ are its generic first integrals, one has
\begin{equation} \label{phis}
{\cal L}_v \, d\phi_{kl}\equiv d (\dot \phi_{kl}) =0 \, .
\end{equation}
On the other hand, the functions $\psi_k$ are particular integrals of the flow.
Then for the chosen basis of such functions and for any $k$,
$$
{\cal L}_v d \psi_k = d (\dot \psi_k)
= \sum_{s=1}^m (\varkappa_{ks} d \psi_s +\psi_s d \varkappa_{ks})  \quad
\mbox{with $\varkappa_{kk}=0$ on ${\cal O}_r$}\, .
$$
Hence,  at points of ${\cal O}_r$ we have
\begin{equation} \label{psis}
d\psi_1\wedge\cdots\wedge ({\cal L}_{v} \, d \psi_k) \wedge
\cdots\wedge  d\psi_m \prod_{1\le k\le l\le r} \wedge \, d \,\phi_{kl} \equiv 0\, .
\end{equation}
Now, combining relations (\ref{IM3}) and (\ref{forms}) with $\xi=$const, we get
$$
{\cal L}_v \Theta \equiv {\cal L}_v
\bigg( 1/\sqrt{ \det ({\mathcal I}|_{{\cal D}_r}) }\;
d\psi_1\wedge\cdots\wedge d\psi_m \wedge \varSigma^N
\prod_{1\le k\le l\le r} \wedge \, d\phi_{kl} \bigg)=0\, .
$$
This, together with (\ref{phis}), (\ref{psis}), implies that
$
{\cal L}_v ( 1/\sqrt{ \det ({\mathcal I}|_{{\cal D}_r})  } \, \varSigma^N)=0
$;
therefore the volume form under the Lie derivative is an integral invariant.
In view of the symplectomorphism between ${\cal O}_r$ and $T^* V(r,n)$,
we can replace the  volume form $\varSigma^N$ by $\varOmega^N$, which
yields the statement of the theorem.
\medskip

Reducibility of the system (\ref{dot_P})  to the Hamiltonian form via a
time rescaling for an arbitrary rank $r>1$ and an arbitrary inertia tensor
$\cal I$ is an open problem.

\section{Rank 1 case and integrable geodesic flow on $S^{n-1}$}

Now we concentrate on the case $r=1$ given by the original
condition (\ref{44}) and again assume that the inertia tensor has the
form (\ref{4.5}). The variety  $V(1,n)$ can be realized as the
unit sphere $S^{n-1}$ in $\mathbb{\ R}^n=(q_1,\dots,q_n)$,
\begin{equation*}
S^{n-1}=\{q_1^2+\cdots+q_n^2=1\},
\end{equation*}
where we set $q=e_1$, and the momentum map (\ref{4.3}) is reduced to
\begin{equation}
\omega=\Phi(q,\dot q)=q\wedge \dot q. \label{44.3}
\end{equation}
Therefore, for the solution $(e_1(t), \,\omega(t)=e_1(t) \wedge \dot e_1(t))$ of
the system (\ref{4.1c}) (\ref{4.2a}), $q(t)=e_1(t)$ is a motion of the reduced system on
the sphere $S^{n-1}$.

For the analysis of the reduced system we can use Theorems \ref{rest_St}, \ref{mes}
of the previous section. However, for our purposes
we shall use the reduction procedure described in Proposition 2.1.

Under the condition  (\ref{4.5}) on the inertia tensor and in view of
(\ref{44.3}),  the reduced Lagrangian $L(q,\dot q)$ and
the right hand side of the Lagrange-d'Alembert equation (\ref{1.2}) take the form
\begin{gather}
L = \frac 1{2\det A} [(A\dot q,\dot q)(Aq,q)- (Aq,\dot q)^2 ] \, ,  \label{redL} \\
 \langle {\cal I} \Phi(q,\dot q), \pr_{g^{-1}\mathfrak{h} g}
 [\Phi(q,\dot q),\Phi(q, \xi)] \rangle  = \frac 1{\det A} \langle A q \wedge A\dot q,
\pr_{g^{-1}\mathfrak{h} g}\xi \wedge \dot q \rangle  \nonumber \\
 = \frac 1{\det A} (\dot q, A \dot q) (Aq, \xi) -  \frac 1{\det A} (\dot q, A q) (A\dot q, \xi)
= \Psi(q,\dot q,\xi)\, . \label{Pi}
\end{gather}
Here we used the relation $\pr_{g^{-1}\mathfrak{h} g}\xi \wedge \dot q = \xi \wedge\dot q$
for any admissible vector $\xi=(\xi_1,\dots,\xi_n)^T \in T_q S^{n-1}$.

As in Section 4, the reduction of the LR system (\ref{nh3.46}) onto  $T^*S^{n-1}$
can explicitly be written in terms
of local coordinates $q_1,\dots, q_{n-1}$ on $S^{n-1}$ and the corresponding momenta.

As an alternative, below we shall keep using the
redundant coordinates $q_i$ and velocities $\dot q_i$, in which
the Lagrange equations have the form
 \begin{gather}
\frac{\partial L}{\partial q_i} - \frac{d}{dt}\frac{\partial
L}{\partial \dot q_i} =\pi_i +\Lambda q_i \, , \qquad i=1,\dots, n,
 \label{43} \\
\pi_i= \frac{\partial\Psi}{\partial \xi_i}
=\frac 1{\det A}  (\dot q, A \dot q) A_i q_i - \frac 1{\det A}
 (\dot q, A q) A_i \dot q_i, \nonumber
\end{gather}
where $\Lambda$ is a Lagrange multiplier. We now want to represent
the reduced LR system on $T^*S^{n-1}$ as a restriction of a system on
the Euclidean space ${\mathbb R}^{2n}=\{q, p\}$.
Note that $L(q,\dot q)$ is degenerate in the redundant velocities $\dot q$, hence
they cannot be expressed uniquely in terms of the redundant momenta
\begin{equation} \label{big_p}
p_i= \frac{\partial L}{\partial\dot q_i}
\equiv \frac 1{\det A} (q, A q) A_i \dot q_i -\frac 1{\det A} (\dot q, A q) A_i q_i \, .
\end{equation}
In this case one can apply the Dirac formalism for Hamiltonian systems with
constraints in the phase space  (see, e.g., \cite{Dirac, AKN, Moser_var}).
Namely, from (\ref{big_p}) we find that $(q,p)=0$,
hence the cotangent bundle $T^*S^{n-1}$ is realized as a subvariety of
${\mathbb R}^{2n}=(q,p)$ defined by the constraints
$$
\phi_1\equiv (q,q)=1, \quad \phi_2\equiv (q,p)=0.
$$
Under these conditions, relations (\ref{big_p}) can be uniquely inverted to yield
\begin{equation} \label{dot_q}
\dot q= \frac {\det A}{(q,Aq)} \left[ A^{-1}p - (p,A^{-1}q) q\right]\, .
\end{equation}
Next, we note that $\partial L/\partial q_i=\pi_i$. Then, from (\ref{43})
we obtain $\dot p=-\Lambda q$, and from the condition $(\dot q,p)+(q,\dot p)=0$,
\begin{equation} \label{dot_p}
\dot p=-\Lambda q, \qquad \Lambda=\det A \frac {(p,A^{-1}p) - (p,q) (q,A^{-1}p)} {(q,Aq)}\,  .
\end{equation}

The system (\ref{dot_q}), (\ref{dot_p}) on $T^* S^{n-1}$ coincides with the restriction of
the following system on  ${\mathbb R}^{2n}=\{q, p\}$:
\begin{gather*}
\dot q_i=\{q_i, \hat H \}_*\, , \quad \dot p_i=\{p_i, \hat H \}_*  -\hat \pi_i\, , \\
\hat\pi_i(q,p) =\pi_i (q,\dot q(q,p)), \quad \hat H=\frac 12 \det A \frac{ (p, A^{-1}p)}{(q,Aq)}.
\end{gather*}
Here $\{F, G\}_*$ denotes the Dirac  bracket on ${\mathbb R}^{2n}$,
$$
\{F, G\}_* =\{F, G\}
+\frac{ \{F,\phi_1 \}\{G,\phi_2\}- \{F,\phi_2\} \{G,\phi_1\} }
{ \{\phi_1, \phi_2 \} }
$$
and $\{\cdot ,\cdot\}$ is the standard Poisson bracket on ${\mathbb R}^{2n}$.
The latter system has the explicit vector form
\begin{equation} \label{extended}
\begin{aligned}
\dot q & = \frac {\det A}{(q,Aq)} \left[ A^{-1}p - \frac{(p,A^{-1}q)}{(q,q)} q\right]\, ,  \\
\dot p & = - \det A \frac {(p,A^{-1}p) (q,q)- (p,q) (q,A^{-1}p)} {(q,Aq)(q,q)^2}\,q\,  .
\end{aligned}
\end{equation}
The bracket $\{\cdot ,\cdot\}_*$ is degenerate and $\phi_1, \phi_2$ are its Casimir functions.

Notice that from (\ref{44.3}) and  (\ref{dot_q}) we get
$$
\omega= q\wedge  \frac {\det A}{(q,Aq)} \left[ A^{-1}p
- \frac {(p,A^{-1}q)}{(q,q)} q\right]\, .
$$
Then equations (\ref{extended}) can also be obtained directly from
Theorem 5.4 by setting $r=1$, ${\mathcal X}=q, {\mathcal P}=p$.

Finally, from Theorems 5.2 and 5.5 we get the following corollary.
\begin{cor} \label{redLR}
The reduced LR system  (\ref{dot_q}), (\ref{dot_p})  on $T^*S^{n-1}$ possesses an
invariant measure
$$
(Aq,q)^{-(n-2)/2}\, \sigma, \qquad \sigma=\varOmega^{n-1}\, ,
$$
where $\sigma$ is the  volume $2(n-1)$-form and $\varOmega$ is the restriction of the
canonical symplectic form $dp_1\wedge dq_1+\cdots+dp_n\wedge dq_n$ onto
$T^* S^{n-1}$.
\end{cor}

In particular, for the reduction of the Veselova LR system (\ref{3.1}) onto $T^* S^2$,
the density of its invariant measure is proportional to $1/\sqrt{(q, Aq)}$,
as was already claimed in Section 4.
\medskip

\paragraph{Reducibility.}
As follows from Corollary \ref{redLR}, item 1) of Theorem \ref{Th_reduce}, and the
fact that the dimension of the reduced configuration manifold equals $n-1$,
if our reduced LR system on $T^* S^{n-1}$ were transformable to a Hamiltonian form by
a time reparameterization, then the corresponding reducing multiplier $\N$ should be
proportional to $1/\sqrt{(q, Aq)}$.

Although Chaplygin's reducibility theorem does not admit a straightforward
multidimensional generalization, i.e.,  item 1) of Theorem \ref{Th_reduce} cannot be inverted,
remarkably, for our reduced LR system on $T^* S^{n-1}$ the inverse statement becomes applicable.

\begin{thm} \label{main}
\begin{description}
\item{1).} Under the time substitution $d\tau =\sqrt{\det A/ (Aq,q) }\, dt$
and an appropriate change of momenta, the reduced LR system (\ref{43}) or
(\ref{dot_q}), (\ref{dot_p}),
becomes a Hamiltonian system describing a geodesic flow on $S^{n-1}$ with the following
Lagrangian obtained from (\ref{redL})
\begin{equation} \label{L^*}
L^{\ast }(q,d q/d\tau )=\frac{1}{2} ( q, Aq)^{-1}
\left[ \bigg(A \frac{d\, q}{d\tau}, \frac{d\, q}{d\tau} \bigg)(Aq,q)
- \bigg(Aq, \frac{d\, q}{d\tau}\bigg)^2 \right]\, .
\end{equation}

\item{2).} The latter system is algebraic completely integrable
for any dimension $n$. In the spheroconic coordinates
$\lambda_{1}, \dots,\lambda _{n-1}$ on $S^{n-1}$\ such that
\begin{equation} \label{S_n}
q_{i}^{2}=\frac{\left( I_{i}-\lambda _{1}\right) \cdots
\left( I_{i}-\lambda_{n-1}\right) }{\prod_{j\ne i} \left( I_{i} -I_{j}\right) },
\qquad I_i=A^{-1}_i \, ,
\end{equation}
the Lagrangian $L^*(q, dq/d\tau )$ takes the St\"ackel form and the evolution of $\lambda_k$
is described by the Abel--Jacobi quadratures
\begin{gather} \label{quad_N}
\frac{\lambda _{1}^{k-1}d\lambda _{1}}{2\sqrt{R\left( \lambda _{1}\right) }}
+\cdots +\frac{\lambda _{n-1}^{k-1}d\lambda _{n-1} }{2\sqrt{R\left( \lambda
_{n-1}\right) }}=\delta _{k,n-1}\, \sqrt{2h}\, d\tau,  \\
k=1, \cdots ,n-1, \nonumber
\end{gather}
where
\begin{equation} \label{Rad}
R(\lambda) =- ( \lambda -I_{1}) \cdots (\lambda-I_{n})
\lambda (\lambda-c_{2})\cdots (\lambda-c_{n-1}),
\end{equation}
$h=L^*$ being the energy constant and $c_{2},\cdots,c_{n-1}$ being other constants
of motion (we set $c_1=0$).
For generic values of these constants the corresponding invariant manifolds
are $(n-1)$-dimensional tori.
\end{description}
\end{thm}

\medskip

We start with the proof of item 2) of Theorem \ref{main}, which is quite standard.
Using the Jacobi identities,
\begin{equation*}
\mbox{for any distinct $\rho_1,\dots, \rho_N$}, \quad
\sum_{s=1}^N \frac{\rho^m }{\prod_{l\ne s} (\rho_l-\rho_s)}
=\bigg\lbrace \begin{aligned} 0 & \quad \mbox{for  } 0 \le  m <N-1, \\
1 & \quad \mbox{for  }  m = N-1, \end{aligned}
\end{equation*}
 in the spheroconic coordinates we have
\begin{align}
\bigg(A \frac{d\, q}{d\tau}, \frac{d\, q}{d\tau} \bigg) & (Aq,q)
- \bigg(Aq, \frac{d\, q}{d\tau}\bigg)^2 \\
& = \frac 14 \frac{\lambda _{1} \cdots \lambda_{n-1}}
{I_{1}\cdots I_{n}} \sum\limits_{k=1}^{n-1}\frac{\prod_{s\neq k}\left(
\lambda_{k}-\lambda _{s}\right) } {\left( \lambda _{k}-I_{1}\right) \cdots
\left( \lambda _{k}-I_{n}\right) \lambda _{k}} \left( \frac{d}{d\tau } \lambda _{k}\right)^{2},
\nonumber \\
 ( A q,q ) & \equiv ( I^{-1}q,q )  =\frac{\lambda _{1} \cdots
\lambda_{n-1} } {I_{1} \cdots I_{n}} \, .
\label{prod_n}
\end{align}
Then the reduced Lagrangian $L^* (q,dq/d\tau )$ in (\ref{L^*}) takes form
\begin{equation*}
L^{\ast }  =\frac{1}{8}\sum\limits_{k=1}^{n-1}\frac{\prod_{s\neq
k}\left( \lambda _{k}-\lambda _{s}\right) }{\left( \lambda _{k}-I_{1}\right)
\cdots \left( \lambda _{k}-I_{n}\right) \lambda _{k}}\left( \frac{d}{d\tau }
\lambda _{k}\right) ^{2}.
\end{equation*}
As a result, the corresponding Hamiltonian written in terms of
$$
\lambda_k, \quad \mu_k=\frac{\partial L^*}{\partial (d\lambda_k/d\tau)}
$$
is of St\"ackel type (see e.g., \cite{AKN}),
which leads to the quadratures (\ref{quad_N}) and proves the integrability of the system.

The proof of item 1) of Theorem \ref{main} is based on a relation between the reduced LR
system and the Neumann system on $S^{n-1}$ and will be given at the end of this section.
\medskip

\paragraph{Reduction to the Neumann system.}
It appears that Theorem \ref{LR->N3} relating the Veselova LR
system and the classical Neumann  system has the following multidimensional
generalization. Introduce another new time $\tau_1$ by formula
\begin{equation}
d\tau_1=\hat \mu^{-1} dt, \qquad
\hat \mu^{-1}=\sqrt{\det A \frac{ \langle q\wedge\dot q, {\mathcal I} (q\wedge\dot q) \rangle }
 {(Aq,q)}}\, dt,  \label{4.6}
\end{equation}
and let $^{\prime}$ denotes the derivation in the new time.

\begin{thm}  \label{LR->Nn}  Under the time substitution (\ref{4.6}),
the solutions $q(t)$ of the reduced multidimensional Veselova system on $S^{n-1}$
transform to solutions of the integrable Neumann problem with
potential $U(q)=\frac12 (A^{-1} q,q)$,
\begin{equation}
q^{\prime\prime}= -\frac{1}{A} q + \lambda q,  \qquad q^{\prime}=\frac{dq}{d\tau_1}\, ,
\label{4.7}
\end{equation}
corresponding to the zero value of the integral
\begin{equation}
F_0 (q,q^{\prime})=\langle Aq^{\prime},q^{\prime}\rangle \langle Aq,q\rangle-
\langle Aq, q^{\prime}\rangle^2-\langle A q,q\rangle  \label{4.8}
\end{equation}
and vise versa.
\end{thm}

The proof we shall present here is similar to that of "three-dimensional" Theorem 4.1,
which was given in \cite{VeVe2}.

\noindent{\it Proof of Theorem\/} \ref{LR->Nn}.  Let
$\omega=q\wedge\dot q$ as above, and set  $P=q\wedge q'$.  Then the energy
integral of the reduced Veselova system and the integral
(\ref{4.8}) of the Neumann system can be written as
\begin{equation*}
E(q,\dot q)=\frac12 \langle {\mathcal I} \omega,\omega \rangle, \quad
F(q,q^{\prime})=\det{A} \langle {\mathcal I} P,P\rangle - (Aq,q).
\end{equation*}

The time reparameterization (\ref{4.6}) induces a bijection between invariant
submanifolds ${\cal E}_h=\{E=h\}\subset TS^{n-1}\{q,\dot q\}$ and ${\cal F}_0=\{F=0\}\subset
TS^{n-1}\{q, q^{\prime}\}$. Indeed, on ${\cal E}_h$ we have
\begin{equation}
dt =\mu_h d\tau_1\, , \qquad \mu_h^{-1}= \sqrt{\frac{2h\det A}{(Aq,q)}}\, .  \label{4.9}
\end{equation}
Then the point $(q,\dot q)\in {\cal E}_h$ corresponds to
$(q,q^{\prime})$, $q^{\prime}=\mu_h \dot q$, and the equation
$\langle I\omega,\omega\rangle/2=h$
corresponds to the relation
\begin{equation*}
\frac1{2\mu_h^2} \langle {\mathcal I} P,P\rangle \equiv \frac12\frac{2h\det{A}}
{(Aq, q)} \langle P, {\mathcal I} P\rangle=h.
\end{equation*}
Therefore $F=\det{A}\langle \mathcal I P, P\rangle -(A q,q)=0$, and
$(q,q^{\prime})\in {\cal F}_0$.

Next, note that equations (\ref{4.2}) with $\omega=q\wedge\dot q$ are equivalent to
\begin{equation}
(\mathcal I\dot\omega \cdot q)\wedge q+ (\mathcal I\omega\cdot \dot q)\wedge q=0 \, .
\label{4.10}
\end{equation}

In view of time reparameterization (\ref{4.9}) we have that $P=\mu_h \omega$ and
\begin{equation}
\frac{dP}{d\tau}=\frac{dP}{dt}\mu_h=\frac{d}{dt}(\mu_h \omega)\mu_h
=\mu_h^2 \frac{d\omega}{dt}+\frac12\frac{d}{dt}(\mu_h^2)\omega \, .
\label{4.12}
\end{equation}
Now we apply the inertia operator (\ref{4.5}) to both sides of this relation, then multiply the
result by the vector $q$, and finally take the wedge product with $q$. As a result,
due to (\ref{4.9}), we get
\begin{equation}
({\mathcal I} P^{\prime}\cdot q)\wedge q
 =\frac{(Aq,q)}{2h\det{A}} (\mathcal I\dot\omega \cdot q)\wedge q
+ \frac{(Aq,\dot q)}{2h\det A} (\mathcal I\omega\cdot q)\wedge q \, .
\label{4.13}
\end{equation}
Using  (\ref{4.10}), we transform (\ref{4.13}) into
\begin{equation}
2h\det A \, ({\mathcal I} P^{\prime}\cdot q)\wedge q
= - (Aq,q) (\mathcal I\omega \cdot \dot q)\wedge q +(Aq,\dot q) (\mathcal I\omega\cdot q)\wedge q .
\label{4.14}
\end{equation}
The right hand side of (\ref{4.14}) is of the form $\Xi\wedge q$, where
\begin{eqnarray}
\Xi &=& (Aq,\dot q){\mathcal I}\omega\cdot q-(Aq,q) {\mathcal I} \omega
\cdot\dot q  \notag \\
&=& \frac{1}{\det A} (Aq,\dot q) (Aq\otimes A\dot q-A\dot q\otimes Aq)\cdot q  \notag  \\
&&-\frac{1}{\det A} (Aq,\dot q)(Aq\otimes A\dot q-A\dot q\otimes Aq)\cdot \dot q
=-2h Aq . \label{4.15}
\end{eqnarray}
For the last equality in (\ref{4.15}) we used the identity
\begin{equation*}
2h=\langle {\cal I} (q\wedge \dot q), q\wedge\dot q\rangle = \frac{1}{\det A} \langle
Aq\wedge A\dot q,q\wedge\dot q\rangle
= \frac{1}{\det A} (Aq,q) (A\dot q,\dot q)- (Aq,\dot q)^2.
\end{equation*}
Hence,  (\ref{4.14}) and (\ref{4.15})  yield
\begin{equation}
({\mathcal I} P^{\prime}\cdot q)\wedge q=\frac{1}{\det A} q\wedge Aq, \qquad
P=q\wedge q^{\prime}.  \label{4.11}
\end{equation}
In view of the constraint $(q,q)=1$, this is equivalent to equations (\ref{4.7}).

Thus we proved that if $q(t)$ is a solution of the reduced multidimensional Veselova
system laying on ${\cal E}_h$, i.e., $q(t)$ satisfies (\ref{4.10}), then
$q(\tau_1)$ is a solution of the Neumann system (\ref{4.7}) laying on ${\cal F}_0$.

Conversely, starting from (\ref{4.11}) and repeating calculations in the inverse direction, we
arrive at  (\ref{4.10}). The theorem is proved.
\medskip

It is known (see e.g., \cite{Knorr1, Moser_var, Neum}) that the Neumann system on $S^{n-1}$
possesses the following family of quadratic first integrals:
\begin{equation} \label{family1}
{\cal F}(\lambda) = \sum_{1\le i<j\le n} \frac{P_{ij}^2} {(\lambda-I_i)(\lambda-I_j)}+
\sum_{i=1}^n \frac{q_i^2}{\lambda-I_i},
\end{equation}
and that the evolution of the spheroconic coordinates $\lambda_k$ defined by (\ref{S_n})
is described by equations
\begin{equation} \label{quad_neumann}
\frac{\lambda _{1}^{k-1}d\lambda _{1}}{2\sqrt{{\cal R}( \lambda_{1}) }}
+\cdots +\frac{\lambda _{n-1}^{k-1}d\lambda _{n-1} }{2\sqrt{{\cal R}
( \lambda_{n-1}) }}=\delta _{k,n-1}\, d\tau_1,  \qquad k=1, \cdots ,n-1,
\end{equation}
where ${\cal R}(\lambda)$ is the following polynomial of degree $2n-1$:
$$
{\cal R}= - \Phi^2(\lambda) {\cal F}(\lambda), \qquad
\Phi(\lambda)=( \lambda -I_{1}) \cdots ( \lambda-I_{n})\, .
$$
Next, as follows from (\ref{family1}),
for the trajectories $q(\tau_1)$ corresponding to the zero value of the integral (\ref{4.8}),
we have ${\cal F}(0)=0$. Hence in this case, the polynomial ${\cal R}(\lambda)$
has the same form as (\ref{Rad}), that is
\begin{equation} \label{calR}
{\cal R}(\lambda)= - ( \lambda -I_{1}) \cdots (\lambda-I_{n}) \lambda
(\lambda-c_{2})\cdots (\lambda-c_{n-1}) \, .
\end{equation}

 Now, comparing equations (\ref{quad_neumann}) with
the quadratures (\ref{quad_N}), we arrive at the following proposition.

\begin{prop} \label{N->Geo} Under the time rescaling $d\tau_1= \sqrt{2h} \, d\tau$
the solution $q(\tau_1)$
 of the Neumann system (\ref{4.7}) lying on  ${\cal F}_0=\{F_0=0\}$ transforms to a solution
$q(\tau)$ of the geodesic flow on $S^{n-1}$ described by the Lagrangian $L^*$ in (\ref{L^*})
and having the energy constant $h$, and vise versa. 
\end{prop}

Now combining Theorem \ref{LR->Nn}  and Proposition \ref{N->Geo},  we finally obtain the
proof of item 1) of Theorem \ref{main}.
\medskip

\section{Reconstructed motion on the distribution $D$}

Here we study the integrability of the original (unreduced) LR system on the right-invariant
distribution $D \subset TSO(n)$
of dimension $(n-1)+n(n-1)/2$, which is specified by constraints (\ref{44})
and the left-invariant metric defined by (\ref{4.5}).

In the Hamiltonian case, the integrability of the reduced system generally implies
 a non-commutative integrability of the original system, namely
its phase space is foliated by invariant isotropic tori with quasi-periodic
dynamics. In our nonholonomic case one has to solve the reconstruction problem:
find all trajectories $(g(t),\dot g(t))$ in $D$ that under the $SO(n-1)$--reduction
$\pi: D\to TS^{n-1}$ are projected to the given trajectory $(q(t),\dot q(t))$ in $TS^{n-1}$.
(In particular, for the Fedorov--Kozlov integrable case
of the multidimensional nonholonomic Suslov problem,
the reconstruction problem was studied in \cite{Bl_Z, Bl_Z1}.)

Since  $SO(n-1)$ is a symmetry group of the LR system on $D$, and the reduced motion
on $TS^{n-1}$ occurs on $(n-1)$--dimensional generic invariant tori,
it is natural to expect that the reconstructed motion $(g(t),\dot g(t))$ is
quasi-periodic  over $(\rho+n-1)$--dimensional tori, where $\rho$ does not exceed
the dimension of the maximal commutative subgroup of $SO(n-1)$, that is
$\rho \le \mbox{rank }\; SO(n-1)=\left[\frac{n-1}{2}\right]$ (see \cite{He}).

As we shall see below, for our case this is not true.
In fact, the relation between the reduced LR system and the Neumann system
described by Theorem \ref{LR->Nn} enables us to reconstruct the motion on $D$ exactly.
For this purpose we also shall make use of the remarkable correspondence between the
Neumann system and the geodesic flow on a quadric.
Namely, consider a family of $(n-1)$-dimensional confocal quadrics in
${\mathbb R}^n=(X_1,\dots, X_n)$,
\begin{equation}
Q(\alpha)= \bigg\{ \frac{X^2_1} {\alpha -A_1} +\cdots+\frac{X^2_n}{\alpha - A_n}=-1 \bigg\}\, ,
\quad \alpha\in {\mathbb R} \, ,
\end{equation}
where $A_1,\dots, A_n$ are distinct numbers.

\begin{thm} \textup{(\cite{Knorr1})}.\label{Kno}
Let $ X(s)$ be a geodesic on the quadric $Q(0)$, $s$ being a
natural parameter. Then under the time rescaling
\begin{equation}
ds =\sqrt{\frac {(dX/ds, A^{-1} dX/ds)}{(X, A^{-2} X) }}\,  d
\tau_1 \label{knorrer}
\end{equation}
the unit normal vector $q(\tau_1)= A^{-1} X /| A^{-1} X|$ at the point $X\in Q$ is a
solution to the Neumann system (\ref{4.7}) corresponding to the zero
value of the integral $ F_0 (q,q^{\prime})$ in  (\ref{4.7}) and vise versa.
\end{thm}

Next, it is well known that the problem of geodesics on a  quadric $Q(0)$ is completely integrable
and that the qualitative behavior of the geodesics is described
by the {\it Chasles theorem\/} (see e.g., \cite{Knorr1, Moser_var}): the tangent line
$$
\ell_s=\{X(s)+\sigma\, dX/ds \mid \sigma \in{\mathbb R} \}
$$
of a geodesic $X(s)$ on $Q(0)$ is also tangent to a fixed set of confocal quadrics
$Q(\alpha_2),\dots, Q(\alpha_{n-1})\subset {\mathbb R}^n$, where
$\alpha_2,\dots, \alpha_{n-1}$ are parameters
playing the role of constants of motion (we set $\alpha_1=0$). Now let ${\mathfrak n}_k$ be
the normal vector of the quadric $Q(\alpha_k)$ at the contact point
${\mathfrak p}_k=\ell \cap Q(\alpha_k)$. Then another classical theorem says
that the normal vectors ${\mathfrak n}_1,\dots, {\mathfrak n}_{n-1}$,
together with the unit tangent vector $\gamma=dX/ds$,
form an orthogonal basis in $\mathbb{R}^{n}$.

The integrability of the geodesic flow is also related to the following properties found
by Moser in \cite{Moser_var}.
\begin{prop} \label{Moser1}
\begin{description} \item{1).} Let $x$ be the position vector of any point on the line
$\ell_s$, which is tangent to the geodesic $X(s)$. Then in the new
parametrization $s_1$ such that $ds=- (X, A^{-2} X) \,ds_1$ the
evolution of the line  is described by Lax-type equation in $n\times n$
matrix form
\begin{gather} \frac d{ds_1} {\mathcal L}= [{\mathcal L}, {\mathcal B}],
\qquad  {\mathcal L} = \Pi_ \gamma (A- x\otimes x)\Pi_\gamma, \\
{\mathcal B} = A^{-1} x\otimes A^{-1} \gamma - A^{-1} \gamma\otimes A^{-1} x\, ,
\label{calB}
\end{gather}
where $\Pi_\gamma=Id- (\gamma,\gamma)^{-1}  \gamma \otimes \gamma$ is the projection onto
the orthogonal complement of $\gamma$ in ${\mathbb R}^n$.
\item{2).} The conserved eigenvalues of ${\mathcal L}$ are given by the parameters
$\alpha_1=0, \alpha_2,\dots, \alpha_{n-1}$ of the confocal quadrics and by an extra zero.
The corresponding eigenvectors of ${\mathcal L}$ are parallel to the normal vectors
${\mathfrak n}_1=q,\dots, {\mathfrak n}_{n-1}$, and to $\gamma$.
\end{description}
\end{prop}

Now we are ready to describe generic solutions of the original LR
system on $D\subset T SO(n)$. Let  $q(\tau_1)$ be the solution of
the Neumann system (\ref{4.7}) with $F_0 (q,q^{\prime})=0$, which
is associated to a solution $(q(t), p(t))$ of the reduced LR
system as described by Theorem \ref{LR->Nn}. Let
\begin{equation} \label{Gauss}
X=(q,Aq)^{-1/2} A q(s), \quad {\mathfrak n}_1=q(s),\dots,
{\mathfrak n}_{n-1}(s), \quad \gamma(s)=\frac{dX}{ds}
\end{equation}
be the corresponding geodesic on $Q(0)$ in the parametrization $s$ given by
(\ref{knorrer}) and the unit eigenvectors of $\cal L$ associated to this geodesic.
Note that according to (\ref{4.6}) and
(\ref{knorrer}), one can treat $s$ as a known function of the original time $t$. Then
the following reconstruction theorem holds.

\begin{thm} \label{recon0} For a solution $(q(t), \dot q(t))$ of the reduced LR system on
$T\, S^{n-1}$, a solution $g(t)$ of the original system on the
distribution $D\subset T\,SO(n)$ is given by the orthogonal frame formed by the
unit vectors
$$
e_1=q(t),\; e_2= {\mathfrak n}_2(s(t)),\; \dots,\; e_{n-1}={\mathfrak n}_{n-1}(s(t)),
\quad e_n=\gamma(s(t)).
$$
The other solutions $(g(t), \dot g(t))$ that are projected onto
the same trajectory $(q(t), \dot q(t))$  have the same $e_1$,
while the rest of the frame is obtained by the orthogonal
transformations
\begin{equation} \label{orth}
\big(e_2 (t) \cdots e_n(t) \big)
=\big( {\mathfrak n}_2(t) \cdots {\mathfrak n}_{n-1}(t)\,  \gamma(t)\big )\,{\mathfrak R} \, ,
\end{equation}
where ${\mathfrak R}$ is a constant matrix ranging over the group $SO(n-1)$.
\end{thm}

From Theorems \ref{recon0} and the integrability
properties of the Neumann system on $T^* S^{n-1}$ (Theorem \ref{LR->Nn}), we conclude that
the phase space $D\subset T\,SO(n)$ of the multidimensional
Veselova LR system with the left-invariant metric defined by
(\ref{4.5}) is almost everywhere foliated by $(n-1)$-dimensional
invariant tori, on which the motion is straight-line but not
uniform.

This also implies that, apart from the pull-back of the
$n-1$ integrals of the Neumann system,  the LR system possesses
$(n-1)(n-2)/2$ generally independent integrals on $D$.
In particular, as follows from the nonholonomic momentum equations (\ref{mom_eq1})
with ${\cal W}_k=e_1\wedge e_k$ and the special form of the inertia tensor,
the LR system on $D$ has linear integrals
$$
l_k= \langle {\mathcal M}, e_1\wedge e_k \rangle \equiv
\frac 1{\det A} \langle A\omega A, e_1\wedge e_k \rangle , \qquad k=2,\dots, n,
$$
of which $n-2$ ones are independent, since $l_2^2+\cdots+l_n^2=p^2$.
\medskip

\noindent{\it Proof of Theorem\/} \ref{recon0}. As follows from
Proposition \ref{Moser1}, for the geodesic motion on $Q(0)$, the
unit normal vectors ${\mathfrak n}_1,\dots,{\mathfrak n}_{n-1}$
and $\gamma$ satisfy the equations
\begin{gather*}
\frac d{ds_1} {\mathfrak n}_k =
- {\mathcal B}\, {\mathfrak n}_k, \quad \frac d{ds_1} {\gamma}
=- {\mathcal B}\, {\gamma}, \qquad k=1,\dots, n-1, \\
 ds=-\nu\, ds_1, \qquad \nu=(X, A^{-2} X)\, ,
\end{gather*}
which can be regarded as kinematic (Poisson) equations with the
``angular velocity'' matrix ${\mathcal B}$.

Next, from Theorem \ref{Kno} we have
\begin{equation} \label{X,X}
X=\sqrt{\nu}\, Aq \quad \mbox{and} \quad \gamma\equiv
\frac{dX}{ds}=\sqrt{\nu}\, A \frac{dq}{ds}+ \frac{d\sqrt{\nu} }{ds} Aq.
\end{equation}
Now let us choose $x=X(s)$ in the expression (\ref{calB}) for $\mathcal B$.
Substituting (\ref{X,X}) into this expression we find that
${\mathcal B}=\nu \, q\wedge dq/ds$.
Then, in the original parameterization $s$, the above Poisson equations take the simple
form
$$
d {\mathfrak n}_k/ds = - (q\wedge dq/ds) {\mathfrak n}_k, \quad
d{\gamma}/ds=  - (q\wedge dq/ds) {\gamma}\, .
$$
Changing here the time parameter $s$ to $t$ and taking into account relation
(\ref{44.3}), we finally obtain
\begin{equation} \label{norms}
\dot {\mathfrak n}_k = - \omega (t) \,{\mathfrak n}_k, \quad k=1,\dots,n-1, \quad
\dot{\gamma}= - \omega(t) {\gamma},
\end{equation}
where $\omega\in {\mathcal D}\subset so(n)$ is the admissible angular velocity of the
$n$-dimensional body. This implies that the orthogonal frame
$({\mathfrak n}_1(t),\dots,{\mathfrak n}_{n-1}(t),\gamma(t))$
gives a solution of the LR system on $D$.

Note that the vectors of the frames that are obtained by the orthogonal transformations
(\ref{orth}) also satisfy the Poisson equations (\ref{norms}) and therefore also give
solutions of the LR system on $D$.
Since the fiber of the map $\pi: D\to TS^{n-1}$ is the group $SO(n-1)$,
there are no other solutions
on $D$ that are projected onto the same trajectory $(q(t), \dot q(t))$.
The theorem is proved.
\medskip

\paragraph{The explicit solution for the frame $(q(t),{\mathfrak n}_{k}(t),\gamma(t))$.}
In order to find explicit expressions for the components of
${\mathfrak n}_k$ and $\gamma$,
following Jacobi \cite{Jac} we first introduce ellipsoidal
coordinates $\nu_{1}, \dots,\nu _{n-1}$ on $Q(0)$ according to the formulas
$$
X_{i}^{2}=\frac{A_i ( A_{i}-\nu_1 ) \cdots ( A_{i}-\nu_{n-1} )}
{\prod_{j\ne i} (A_{i} -A_{j} ) }, \qquad i=1,\dots, n .
$$
Matching these with the expressions  (\ref{S_n}) for $q_i$ in terms of the spheroconic
coordinates $\lambda_{1}, \dots,\lambda_{n-1}$ on $S^{n-1}$ and taking into account
(\ref{Gauss}) and (\ref{prod_n}),  we find that,  up to a permutation of indices,
$$
\nu_k=\lambda_k^{-1}, \qquad k=1,\dots,n-1 .
$$
Using this property one can also prove that the {\it nonzero\/} parameters
$\alpha_2,\dots,\alpha_{n-1}$
of the confocal quadrics in the Chasles theorem are just the inverses of the constants
of motion $c_2,\dots,c_{n-1}$ in the invariant polynomial (\ref{calR}),
$$
{\cal R}(\lambda) = - ( \lambda -I_{1}) \cdots (\lambda-I_{n}) \lambda
(\lambda-c_{2})\cdots (\lambda-c_{n-1}) \, .
$$
As a result, making use of the definition of the
vectors ${\mathfrak n}_k$, $\gamma$,
one can express their components in terms of pairs
$(\lambda_{1}, \sqrt{{\cal R}(\lambda_1)}), \dots,(\lambda_{n-1},\sqrt{{\cal R}(\lambda_{n-1})})$
and $c_2,\dots,c_{n-1}$. After some calculations we get
\begin{align}
q_{i} &=\frac{\sqrt{U(\lambda ,I_i)}} {\sqrt{\Psi'(I_i)}}, \qquad I_i=A^{-1}_i \, ,
\nonumber \\
({\mathfrak n}_k)_{i}
&= \frac{\sqrt{U(\lambda ,I_i)}} {\sqrt{\Psi'(I_i)}} \,
\frac{\sqrt{U(\lambda ,c_k)}}{\sqrt{\psi'(c_k)}} \cdot
\sum_{s=1}^{n-1}
\frac{\Xi_s}{(c_k -\lambda_s)(I_i-\lambda_s)}\, , \label{wurzel} \\
\gamma_{i}
&=\frac{\sqrt{U(\lambda ,I_i)}} {\sqrt{\Psi'(I_i)}}\,
\frac{\sqrt{U(\lambda ,0)}}
     {\sqrt{\psi '(0)}} \cdot \sum_{s=1}^{n-1}
\frac{\Xi_s}{ \lambda_s(I_i-\lambda_s)}\, , \nonumber \\
 i &=1,\dots,n, \quad k=2,\dots, n-1 , \nonumber
\end{align}
where
\begin{gather*}
\Psi (r)=(r-I_{1})\cdots(r-I_{n})\, ,  \quad
\psi (r)= r(r-c_{2})\cdots(r-c_{n-1})\, ,    \\
\Psi'(I_i)= \frac d{dr} \Psi(r)|_{r=I_i}, \quad
\psi'(c_k)= \frac d{dr} \Psi(r)|_{r=c_k}, \\
U(\lambda,r)=(r-\lambda_{1})\cdots(r-\lambda_{n-1})\, , \quad
\Xi_{s}= \frac{\sqrt{{\cal R}(\lambda_s)}}
{\prod\limits_{j\ne s}^{n-1}(\lambda_{s}-\lambda_{j})}\, .
\end{gather*}
The evolution of $\lambda$-coordinates in the time $\tau$ is described
by the quadratures (\ref{quad_N}), (\ref{Rad}).

The squares of expressions (\ref{wurzel}) are symmetric algebraic functions of $n-1$ points
$(w_k, \lambda_k)$ on the hyperelliptic curve ${\cal C}=\{w^2 ={\cal R}(\lambda)\}$ of genus $n-1$.
Then, by using the classical algebraic geometrical methods (see, e.g., \cite{Baker1, Cl_Gor}), the
components of  $q, {\mathfrak n}_2,\dots, {\mathfrak n}_{n-1}$, $\gamma$, and
the function $\sqrt{\lambda_1 \cdots \lambda_{n-1}}$ can be represented as quotients of
theta-functions with half-integer theta-characteristics associated to $\cal C$,
whose arguments depend linearly on $\tau_1$.

Finally, the dependence of $t$ in $\tau_1$
is obtained by the integration of (\ref{4.6}), which, in view of (\ref{prod_n}), leads to the
simple quadrature
$$
t= \frac{1}{\sqrt{2h}} \int \sqrt{\lambda_1(\tau_1) \cdots
\lambda_{n-1}(\tau_1)}\, d\tau_1 +\mbox{const} .
$$

\section*{Conclusion}
In this paper we considered  LR systems on compact Lie groups and showed that their
reductions to homogeneous spaces always possess an invariant measure.
We calculated the density of this measure explicitly in case of the Stiefel varieties $V(r,n)=SO(n)/SO(n-r)$.
It turned out that for $r=1$ and the special inertia tensor on $so(n)$,
the reduced flow is transformed to
an integrable geodesic flow on $S^{n-1}$ via the time rescaling prescribed
by the density of the invariant measure and Theorem \ref{Th_reduce}.
Moreover, in this case the unreduced flow on the right-invariant distribution
$D \in T\, SO(n)$ is also integrable.

Such a behavior of a multidimensional nonholonomic system
is exceptional, which may follow from the rich underlying
geometry coming from the Chasles theorem and the Jacobi problem on geodesics on an ellipsoid.
(The latter is closely related to integrability of various problems in mechanics.

In this connection the following questions arise: are there other  inertia tensors of
LR systems on $SO(n)$ for which the  above properties hold and how wide is the class
of such tensors? Can reduced flows on $V(r,n)$, $r>1$ be transformed to the Hamiltonian form
(with respect to the canonical symplectic structure $\varOmega$) in the same manner?
Are the reduction to a Hamiltonian form and integrability still possible in case of
nonhomogeneous right-invariant constraints on $SO(n)$
(similar to what takes place for the classical case $n=3$)?

A part of our analysis can be extended to LR systems on noncompact Lie groups and their
reductions. It would be interesting  to study meaningful examples of such systems.

\subsection*{Acknowledgments}
We thank A.~Bolsinov, A. Ramos, and D. Zenkov for useful discussions during preparation
of the manuscript, as well as J. Orlich for assistance in english language editing.

The first author (Yu.F.) acknowledges the support of the Russian Federation program
of supporting academic groups No. 00-15-96146.
The second author (B.J.) was partially supported by the Serbian Ministry of Science
and Technology, Project 1643 (Geometry and Topology of Manifolds and
Integrable Dynamical Systems).

\end{document}